\documentclass[a4paper,11pt]{article}
\usepackage{latexsym, amsfonts, amsthm,amssymb,amscd,amsmath,makeidx,bbm,color}
\usepackage{graphicx}
\usepackage{titlesec}
\usepackage{eurosym}
\usepackage{float}
\usepackage{caption}
\usepackage{color,setspace}
\usepackage[margin=1in]{geometry}
\usepackage{pdflscape}
\usepackage{subcaption}

\allowdisplaybreaks
\DeclareMathAlphabet{\mathcalligra}{T1}{calligra}{m}{n}

\theoremstyle{plain}
\newtheorem{theorem}{Theorem}[section]

\newtheorem{proposition}[theorem]{Proposition}

\theoremstyle{definition}
\newtheorem{definition}[theorem]{Definition}
\newtheorem{example}{Example}[section]
\newtheorem{remark}[theorem]{Remark}
\newtheorem{assumption}{Assumption}[section]

\numberwithin{equation}{section}
\numberwithin{table}{section}
\numberwithin{figure}{section}
\input{tcilatex}
\begin{document}

\title{Pricing FX Options under Intermediate Currency}
\author{S. Maurer\thanks{%
School of Mathematical Sciences, University of Nottingham, UK;
Simon.Maurer@nottingham.ac.uk} \and T.E. Sharp\thanks{%
Credit Suisse, Zurich, Switzerland} \and M.V. Tretyakov\thanks{%
School of Mathematical Sciences, University of Nottingham, UK;
Michael.Tretyakov@nottingham.ac.uk}}
\maketitle

\begin{abstract}
We suggest an intermediate currency approach that allows us to price options
on all FX markets simultaneously under the same risk-neutral measure which
ensures consistency of FX option prices across all markets. In particular,
it is sufficient to calibrate a model to the volatility smile on the
domestic market as, due to the consistency of pricing formulas, the model
automatically reproduces the correct smile for the inverse pair (the foreign
market). We first consider the case of two currencies and then the
multi-currency setting. We illustrate the intermediate currency approach by
applying it to the Heston and SABR stochastic volatility models, to the
model in which exchange rates are described by an extended skewed normal
distribution, and also to the model-free approach of option pricing.

\noindent \textbf{Keywords}: foreign exchange market, FX option pricing,
foreign-domestic symmetry, multi-currency options, skewed normal
distribution.
\end{abstract}

\section{Introduction\label{sec:intro}}

As it is well known (see e.g. \cite{Bjork, Wystup} and also Section~\ref%
{sec:markets} here), in the case of a foreign exchange (FX) for two
currencies (say, GBP and EUR) no measure is simultaneously risk-neutral for
the market on which GBP is the domestic currency and for the market on which
EUR is the domestic currency. This can be seen as an asymmetry between the
different market views due to the different choice of numeraires. In
practice currency pair conventions are usually used in order to standardize
option price quotations for each specific currency pair \cite{Lipton, Wystup}%
. But this can lead to calibration difficulties. Each of the domestic
markets has its own volatility smile curve for options on the corresponding
foreign currency. Suppose we want to use a stochastic volatility model given
under a risk-neutral measure on the GBP domestic market which we calibrate
to the smile for options on EUR. If we re-write this model for the inverse
pair, i.e., where options on GBP are traded, in a risk-neutral fashion, we
need to calibrate it to the smile on this market as the previously found
parameters of the model typically do not match the smile for the inverse
pair. This is inconvenient. This situation becomes even more complicated in
a multi--currency setting while it is of practical importance to be able to
price options on the global FX market in a consistent fashion. With a large
number $N$ of currencies, the existence of a consistent FX model is not
trivial as a suitable model must preserve relationships between all $N$
currencies and consistency of volatility smiles between all $N(N-1)/2$ cross
pairs.

To address these problems of consistent FX modelling, in \cite%
{doust2007intrinsic} (see also \cite%
{doust2008estimating,doust2012stochastic,de2013smiles}) the concept of
intrinsic currency \cite{doust2007intrinsic,doust2012stochastic} or
artificial currency \cite{de2013smiles} was introduced. The approach of \cite%
{doust2007intrinsic} is based on the idea that each currency has an
`intrinsic value', which is a description of the value of a currency in
relation to other currencies. In the intrinsic currency-valuation framework
of \cite{doust2007intrinsic,doust2012stochastic} one models the $N$
intrinsic values of $N$ currencies rather than modelling the $N-1$ exchange
rates. In \cite{doust2012stochastic} Doust extends his original idea of the
intrinsic currency-valuation framework to a SABR-type model, which allows to
capture the observed volatility smile on the FX market in a multi-currency
setting. On a FX market with $N$ currencies, he describes the market with $N$
intrinsic currency values and chooses one (without loss of generality) as
the valuation currency and its associated risk--neutral measure, which
produces the usual risk--neutral processes for all exchange rates. In \cite%
{de2013smiles} $N$ exchange rates between an artificial currency and $N$
real currencies are modelled under a risk-neutral measure associated with
the artificial currency so that all relationships (in particular, the
inversion property that the exchange rate for a pair of real currencies and
for their inverse satisfy SDEs of a similar form) between $N$ currencies are
satisfied.

Here we explore a very simple but very valuable from the practical angle
idea: find a numeraire with respect to which we can price all FX derivatives
traded on any of the domestic markets simultaneously under the same measure.
This resolves the issue highlighted above: models for different currency
pairs can be calibrated to all smiles in a consistent manner. For instance,
in the case of two currencies, it is sufficient to calibrate a model on e.g.
the GBP domestic market and the smile on the EUR domestic market is
automatically reproduced without any need of additional calibration.

We show that such a numeraire exists via introducing the concept of an
intermediate pseudo-currency. The main difference with \cite%
{doust2007intrinsic,doust2008estimating,doust2012stochastic,de2013smiles} is
that the pseudo-currency is explicitly defined via exchange rates of real
currencies, while in \cite%
{doust2007intrinsic,doust2008estimating,doust2012stochastic,de2013smiles}
exchange rates of real currencies are described via an artificial, not
observed currency. Consequently, we naturally model $N-1$ exchange rates,
not $N$ as in \cite%
{doust2007intrinsic,doust2008estimating,doust2012stochastic,de2013smiles}.
Further, we can use three modelling approaches in a consistent manner. The
first one is the traditional modelling way in Financial Mathematics, where
we start from a stochastic model for $N-1$ exchange rates under a `market'
measure and then we introduce a pseudo-currency market which, as we show,
has a risk-neutral measure. Under this risk-neutral measure (the
intermediate pseudo-currency is used as the numeraire) we can price FX
products on all currency markets simultaneously which guarantees consistency
of volatility smiles and other natural relationships between currencies
(e.g., the foreign-domestic symmetry). This approach allows us to start with
popular stochastic volatility models (e.g., Heston or SABR) written under a
`market' measure and derive the corresponding consistent models on the
pseudo-currency market. Alternatively, in the second approach, from the
start we model exchange rates under a risk-neutral measure or under a
forward measure associated with the pseudo-currency market. The third
approach is model-free (see \cite{Aus11,Fuk12,Aus14,MM18} and references
therein), where we reconstruct a risk-neutral measure or a forward measure
from volatility smiles. We note that the intermediate pseudo-currency in
comparison with the intrinsic currency of \cite{doust2007intrinsic} does not
have a financial interpretation, but our focus here is solely on consistent
calibration and modelling of exchange rates.

The rest of the paper is organized as follows. In Section~\ref{sec:markets}
we recall that there is no measure which is simultaneously risk-neutral for
both domestic and foreign FX markets and also recall the foreign-domestic
symmetry. A convenient numeraire and the associated intermediate
pseudo-currency market are introduced in Section~\ref{sec:intpricing}, where
the corresponding pricing formulas for FX options are also derived. This is
done for clarity of the exposition in the case of a single currency pair. We
extend the intermediate pseudo-currency concept to the multi--currency
setting in Section~\ref{sec:multiN}. In Section~\ref{sec:ill} we illustrate
the concept by first applying it to the Heston model \cite{Hes} and SABR 
\cite{hagan2002managing}. Then, for further illustration, we model the spot
exchange rate using an extended skewed normal distribution. This exchange
rate model is an illustration of how one can describe the observed
fat-tailed distribution of the log exchange rate (compared to the log normal
assumption). The considered extended skewed normal distribution is
constructed by combining one normal and two shifted half-normal distributed
random variables and it allows a flexible control of the tails of the spot
exchange rate distribution. We note that the use of the extended skewed
normal distribution in pricing FX options is somewhat new. Further, we
illustrate our FX option pricing mechanism on the model-free approach. We
provide some calibration examples in Section~\ref{sec:calib}.

\section{Preliminaries \label{sec:markets}}

In this section we recall the well-known fact (see e.g. \cite{Bjork, Wystup}%
) that there is no measure which is simultaneously risk-neutral for both the
domestic and the foreign market. We also state the foreign-domestic
symmetry. For definiteness, in this section and in Section~\ref%
{sec:intpricing}\ we use the EUR-USD and USD-EUR pairs.

Let us recall that the EUR-USD spot exchange rate at time $t$ 
\begin{equation*}
f(t):=S_{\text{\euro }/\$}(t)
\end{equation*}%
is quoted as 
\begin{equation*}
\frac{\text{units of USD}}{\text{one EUR}},
\end{equation*}%
and 
\begin{equation*}
S_{\$/\text{\euro }}(t)=\frac{1}{S_{\text{\euro }/\$}(t)}=\frac{1}{f(t)}.
\end{equation*}%
In currency pairs (e.g. EUR-USD), the first mentioned currency is known as
the foreign (or base) currency, while the second is known as the domestic
currency (or numeraire) \cite{castagna, Wystup}.

Within the standard option pricing setting, we assume that the currency
market under a `market' measure is described by the system: 
\begin{eqnarray}
&&dB_{\$}=r_{\$}(t)B_{\$}dt,  \label{eq:model1} \\
&&dB_{\text{\euro }}=r_{\text{\euro }}(t)B_{\text{\euro }}dt,  \notag \\
&&df=\mu (t)fdt+\sigma (t)fdW(t),  \notag
\end{eqnarray}%
where $B_{\$}(t)$, $B_{\text{\euro }}(t)$ and $r_{\$}(t)$, $r_{\text{\euro }%
}(t)$ are USD and EUR bank accounts with their short interest rates,
respectively; $\sigma (t)>0$ is a volatility, $\mu (t)$ is a drift; and $%
W(t) $ is a standard Wiener process. It is assumed that the coefficients $%
r_{\$}(t)$, $r_{\text{\euro }}(t)$, $\sigma (t)$, and $\mu (t)$ are
stochastic processes adapted to a filtration $\mathcal{F}_{t}$ to which $%
W(t) $ is also adapted (typically, in stochastic volatility models $\mathcal{%
F}_{t}$ is larger than the natural filtration of $W(t)),$ and they have
bounded second moments. We also require that $\sigma (t)$ satisfies
Novikov's condition.

On the USD market, the foreign currency EUR is paid for by USD (the domestic
currency) and the risky asset is 
\begin{equation*}
Y_{\text{\euro }/\$}(t)=S_{\text{\euro }/\$}(t)B_{\text{\euro }}(t),
\end{equation*}%
while on the EUR market the risky asset is 
\begin{equation*}
Y_{\$/\text{\euro }}(t)=S_{\$/\text{\euro }}(t)B_{\$}(t).
\end{equation*}%
Following the classical theory of pricing, we have to find equivalent
(local) martingale measures (EMMs) $Q^{\$}$ and $Q^{\text{\euro }}$ under
which the corresponding discounted risky assets are (local) martingales. By
standard arguments we arrive at the SDEs for $f(t)$ and $g(t):=1/f(t)$
written under the corresponding EMMs: 
\begin{eqnarray}
df &=&(r_{\$}(t)-r_{\text{\euro }}(t))fdt+\sigma (t)fdW^{Q^{\$}}(t),
\label{eq:sdef} \\
dg &=&(r_{\text{\euro }}(t)-r_{\$}(t))gdt-\sigma (t)gdW^{Q^{\text{\euro }%
}}(t),  \label{eq:sdeg}
\end{eqnarray}%
where $W^{Q^{\$}}(t)$ is a standard Wiener process under $Q^{\$}$ and $W^{Q^{%
\text{\euro }}}(t)$ is a standard Wiener process under $Q^{\text{\euro }}$.
We can see (cf. (\ref{eq:model1}) and (\ref{eq:sdef})-(\ref{eq:sdeg})) that
the market prices of risk on the two markets differ: 
\begin{equation*}
\gamma _{\text{\euro }}(t)=\frac{\mu (t)+r_{\text{\euro }}(t)-r_{\$}(t)}{%
\sigma (t)}\neq \frac{\sigma ^{2}(t)-\mu (t)+r_{\$}(t)-r_{\text{\euro }}(t)}{%
-\sigma (t)}=\gamma _{\$}(t)
\end{equation*}%
(recall that $\sigma (t)>0).$ Thus, 
\begin{equation}
Q^{\$}\neq Q^{\text{\euro }},  \label{eq:QQ}
\end{equation}%
i.e., there is no measure which is simultaneously risk-neutral for the EUR
domestic market and for the USD domestic market in this rather general
setting. We remark that (\ref{eq:QQ}) does not lead to an arbitrage
opportunity.

Note that the SDE (\ref{eq:sdef}) for $f$ under the measure $Q^{\text{\euro }%
}$ takes the form 
\begin{equation}
df=(r_{\$}(t)-r_{\text{\euro }}(t)+\sigma ^{2}(t))fdt+\sigma (t)fdW^{Q^{%
\text{\euro }}}.  \label{eq:sdef2}
\end{equation}%
Intuitively, one could think that the drift for the exchange rate $%
g(t)=1/f(t)$ in (\ref{eq:sdeg}) should be the negative of the drift of $f(t)$
under the same measure, i.e. $-(r_{\text{\euro }}(t)-r_{\$}(t))=r_{\$}(t)-r_{%
\text{\euro }}(t)$. However, as we can see in (\ref{eq:sdef2}), this is not
the case. This is related to the phenomenon known as Siegel's paradox \cite%
{siegel1972risk}, which is due to the convexity of the function $1/f$.

Let us also recall \cite{Grab83,Lipton, Wystup,NaF} that under the
no-arbitrage assumption (and other standard conditions like no transaction
costs, etc.), there is the so-called foreign-domestic symmetry for FX
options which we formulate in the following theorem. This symmetry is the
key requirement for a model to be consistent for a currency pair and its
inverse pair (see e.g. \cite%
{doust2007intrinsic,doust2012stochastic,de2013smiles,gnoatto2014affine} and
references therein).

\begin{theorem}
\label{thm:fdsymmetry} Under the no-arbitrage assumption, there is the
following relationship (called \textbf{Foreign-Domestic Symmetry}) for FX
options 
\begin{equation}
C_{\text{\euro }/\$}(0,T,K)=S_{\text{\euro }/\$}(0)\,K\,P_{\$/\text{\euro }%
}\left( 0,T,\frac{1}{K}\right) ,  \label{eq:sym}
\end{equation}%
where $C_{\text{\euro }/\$}(0,T,K)$ is the call option price (in $\$$) at
time $0$ to buy one EUR for $\$K$ at time $T$; $P_{\$/\text{\euro }%
}(0,T,1/K) $ is the put option price (in $\text{\euro }$) at time $0$ to
sell one USD for \euro $\dfrac{1}{K}$ at time $T$.
\end{theorem}

Let us emphasise that the proof of this theorem is solely based on the
no-arbitrage argument, and hence it states a fundamental property of the FX
market. Suppose we take a stochastic volatility model (e.g., a popular model
such as the Heston and SABR) and calibrate it using option data on the USD
market. If we rewrite this model with the obtained parameters for the
inverse pair \euro $/\$,$ then option prices computed by this model on the
EUR market would not match the data on this market and the property (\ref%
{eq:sym}) would not be satisfied, i.e. we would get an arbitrage. Instead,
if we calibrate the inverse pair model again but using option data on the
EUR market, then the property (\ref{eq:sym}) is obviously satisfied, but it
is inconvenient that the model needs to be calibrated twice despite the fact
that the two smiles are consistent with each other due to absence of
arbitrage and the symmetry (\ref{eq:sym}). We note in passing (see e.g. \cite%
{Rollin,de2013smiles}) that for the SABR and Heston models there are
mappings between the parameters obtained for USD-EUR and the parameters of
the inverted world (i.e., EUR-USD), still the parameters are different for
the direct and inverted worlds.

In the next section we find a numeraire allowing to price options on USD and
EUR markets simultaneously after a single calibration. In particular, within
the proposed approach, calibration of a stochastic volatility model using FX
data from one of the domestic markets guarantees replication of volatility
smiles by the model on both domestic markets.

\section{FX option pricing via intermediate pseudo-currency\label%
{sec:intpricing}}

In this section we propose a candidate for numeraire using which options on
USD and EUR markets can be priced simultaneously under the same measure. It
is convenient to cast introduction of such a numeraire using the notion of
an artificial currency, which we call an intermediate pseudo-currency in
this paper to distinguish from the intrinsic currency of \cite%
{doust2007intrinsic,doust2012stochastic} and the artificial currency of \cite%
{de2013smiles}. We note that the intermediate currency market is virtual and
is only used as a proxy to find a suitable numeraire and write down the
corresponding pricing formulas, while calibration is done using the usual FX
data. We start with the definition of the pseudo-currency, then (Section~\ref%
{sec:QX}) we consider pricing under an EMM $Q^{X}$ on the pseudo-market and
(Section~\ref{sec:QXT}) -- under the T-forward measure $Q_{T}^{X}$
equivalent to $Q^{X}$.

\begin{definition}
\label{def:pseudocurrency} Let $S_{\text{\euro }/\$}(t)=f(t)$ be the EUR-USD
exchange rate at time $t$. An \textbf{intermediate pseudo-currency X} is a
currency with exchange rate EUR-X, $S_{\text{\euro }/X}(t)=\sqrt{f(t)},$ and
the exchange rate USD-X, $S_{\$/X}(t)=\frac{1}{\sqrt{f(t)}}$.
\end{definition}

We observe the natural relationship for the intermediate currency%
\begin{equation}
S_{\text{\euro }/X}(t)\cdot \frac{1}{S_{\$/X}(t)}=f(t).  \label{a1}
\end{equation}%
We note the following symmetry: 
\begin{equation*}
S_{\text{\euro }/X}(t)=\sqrt{f(t)}=\frac{1}{\frac{1}{\sqrt{f(t)}}}=\frac{1}{%
S_{\$/X}(t)}=S_{X/\$}(t)
\end{equation*}%
and 
\begin{equation*}
S_{\$/X}(t)=\frac{1}{\sqrt{f(t)}}=\frac{1}{S_{\text{\euro }/X}(t)}=S_{X/%
\text{\euro }}(t).
\end{equation*}

We also introduce the money market account $B_{X}$ for the intermediate
currency X with its respective interest rate $r_{X}(t)$: 
\begin{equation}
dB_{X}=r_{X}(t)B_{X}dt.  \label{a3}
\end{equation}

In the next section we first establish using Girsanov's theorem that for a
sufficiently broad class of models for $f(t)$ there is an EMM $Q^{X}$ on the
pseudo-market (more precisely, we will find a suitable numeraire) and then,
assuming existence of an EMM $Q^{X},$ we write pricing formulas.

\begin{remark}
\label{rem:two} We can introduce $S_{\text{\euro }/X}(t)=f^{\alpha }(t)$
with any $\alpha \in (0,1),$ then $S_{X/\$}(t)=f^{\alpha -1}(t).$ Each
particular $\alpha $ leads to the corresponding numeraire suitable for the
stated purposes (but note that the numeraires associated with the original
USD and EUR markets are not suitable for the set objective as discussed in
Section~\ref{sec:markets}). Arbitrariness of $\alpha $ can potentially be
used for calibration purposes but we do not consider this aspect here. For
clarity and also for the sake of symmetry, we choose to use $\alpha =1/2$ in
this paper.
\end{remark}

\begin{remark}
We do not attach any economic interpretation to the intermediate currency.
Our interest is purely motivated by calibration aspects. We also note that
we model a single exchange rate which is natural, not $2$ rates as in \cite%
{doust2007intrinsic,doust2008estimating,doust2012stochastic,de2013smiles}.
\end{remark}

\subsection{An EMM for the intermediate market\label{sec:QX}}

Consider the virtual market which domestic currency is X. On this market we
have two risky assets: USD paid by X and EUR paid by X: 
\begin{equation}
Y_{\text{\euro }/X}(t)=S_{\text{\euro }/X}(t)B_{\text{\euro }}(t),\ \ \
Y_{\$/X}(t)=S_{\$/X}(t)B_{\$}(t).  \label{a4}
\end{equation}%
Assume that EUR-USD exchange rate $f(t)$ satisfies the model (\ref{eq:model1}%
). Based on (\ref{eq:model1}), we can write the SDEs under market measure
for $Y_{\text{\euro }/X}(t)$ and $Y_{\$/X}(t)$: 
\begin{eqnarray*}
dY_{\text{\euro }/X} &=&\frac{1}{2}\left( \mu (t)+2r_{\text{\euro }}(t)-%
\frac{\sigma ^{2}(t)}{4}\right) Y_{\text{\euro }/X}dt+\frac{\sigma (t)}{2}Y_{%
\text{\euro }/X}dW(t), \\
dY_{\$/X} &=&\frac{1}{2}\left( -\mu (t)+2r_{\$}(t)+\frac{3\sigma ^{2}(t)}{4}%
\right) Y_{\$/X}dt-\frac{\sigma (t)}{2}Y_{\$/X}dW(t).
\end{eqnarray*}%
If we choose the intermediate currency interest rate $r_{X}$ equal to 
\begin{equation}
r_{X}(t)=\frac{r_{\$}(t)+r_{\text{\euro }}(t)}{2}+\frac{\sigma ^{2}(t)}{8},
\label{eq:rxxx}
\end{equation}%
then there is an EMM $Q^{X}$ for the pseudo-currency market with the
following market price of risk $\gamma (t)$: 
\begin{equation}
\gamma (t)=\frac{\mu (t)-\frac{\sigma ^{2}(t)}{2}+r_{\text{\euro }%
}(t)-r_{\$}(t)}{\sigma (t)},  \label{eq:gamX}
\end{equation}%
i.e. 
\begin{eqnarray*}
dY_{\text{\euro }/X} &=&r_{X}(t)Y_{\text{\euro }/X}dt-\frac{\sigma (t)}{2}Y_{%
\text{\euro }/X}dW^{Q^{X}}, \\
dY_{\$/X} &=&r_{X}(t)Y_{\$/X}dt+\frac{\sigma (t)}{2}Y_{\$/X}dW^{Q^{X}},
\end{eqnarray*}%
where $W^{Q^{X}}$ is the standard Wiener process under $Q^{X}$. Thus, we
have shown that the intermediate pseudo-currency market can be arbitrage
free within a sufficiently broad setting. We summarise this result in the
following statement.

\begin{theorem}
\label{prop:QX}Assume that the EUR-USD currency market under a `market'
measure is described by the model (\ref{eq:model1}). Then there is the
unique intermediate currency interest rate $r_{X}(t)$ defined in (\ref%
{eq:rxxx}) and an EMM $Q^{X}$ for the intermediate pseudo-currency market
with the market price of risk $\gamma (t)$ from (\ref{eq:gamX}), i.e., under
(\ref{eq:rxxx}) the market is arbitrage-free.
\end{theorem}

We see from (\ref{eq:rxxx}) that even if the short rates $r_{\$}(t)$ and $r_{%
\text{\euro }}(t)$ are assumed to be constant, the intermediate currency
interest rate $r_{X}(t)$ is non-constant if the volatility $\sigma (t)$ is
time-dependent. In particular, if $\sigma (t)$ is a stochastic process, then
so is the short rate $r_{X}(t)$.

\begin{example}[\textit{An analogue of the Garman-Kohlhagen formula}]
\label{ex:32} Assume that the exchange rate between EUR and USD $f(t)=S_{%
\text{\euro }/\$}(t)$ satisfies the model (\ref{eq:model1}) with constant
coefficients: $\sigma (t)=\sigma $, $r_{\text{\euro }}(t)=r_{\text{\euro }}$
and $r_{\$}(t)=r_{\$}$. Note that in this simplified case (the geometric
Brownian motion case) the intermediate currency interest rate $r_{X}$ is
constant. Analogously, to the standard derivation of the Garman-Kohlhagen
formula, we can find option prices for a pseudo-currency market investor.
For a European floating-strike call option (priced in X) to buy $1$ EUR for $%
\frac{K}{\sqrt{f(T)}}$ X, we have 
\begin{eqnarray}
\mathcal{E}_{\text{\euro }/X}^{C}(0,T,f(0),K,r_{\$},r_{\text{\euro }})
&=&e^{-r_{X}T}\mathbb{E}_{Q^{X}}\left[ \left( \sqrt{f(T)}-\frac{K}{\sqrt{f(T)%
}}\right) _{+}\right]  \label{eq:icall} \\
&=&\sqrt{f(0)}e^{-r_{\text{\euro }}T}N\left( \frac{\log \frac{f(0)}{K}%
+(r_{\$}-r_{\text{\euro }}+\frac{\sigma ^{2}}{2})T}{\sigma \sqrt{T}}\right) 
\notag \\
&&-\frac{K}{\sqrt{f(0)}}e^{-r_{\$}T}N\left( \frac{\log \frac{f(0)}{K}%
+(r_{\$}-r_{\text{\euro }}-\frac{\sigma ^{2}}{2})T}{\sigma \sqrt{T}}\right) .
\notag
\end{eqnarray}%
And, similarly for a European floating-strike put option (priced in X) to
sell $1$ USD for $\frac{\sqrt{f(T)}}{K}$ X we have: 
\begin{eqnarray}
\mathcal{E}_{\$/X}^{P}(0,T,\frac{1}{f(0)},\frac{1}{K},r_{\text{\euro }%
},r_{\$}) &=&e^{-r_{X}T}\mathbb{E}_{Q^{X}}\left[ \left( \frac{\sqrt{f(t)}}{K}%
-\frac{1}{\sqrt{f(t)}}\right) _{+}\right]  \label{eq:iput} \\
&=&\frac{\sqrt{f(0)}}{K}e^{-r_{\$}T}N\left( \frac{\log \frac{f(0)}{K}%
+(r_{\$}-r_{\text{\euro }}+\frac{\sigma ^{2}}{2})T}{\sigma \sqrt{T}}\right) 
\notag \\
&&-\frac{1}{\sqrt{f(0)}}e^{-r_{\text{\euro }}T}N\left( \frac{\log \frac{f(0)%
}{K}+(r_{\$}-r_{\text{\euro }}-\frac{\sigma ^{2}}{2})T}{\sigma \sqrt{T}}%
\right) .  \notag
\end{eqnarray}

From (\ref{eq:icall}) and (\ref{eq:iput}), we can deduce prices for the call 
$C_{\text{\euro }/\$}$ and put $P_{\$/\text{\euro }}$. To this end, we first
observe that the in-the-money payoff of the option $\mathcal{E}_{\text{\euro 
}/X}^{C}$ (priced in X) is equivalent to buying \euro $1$ for $\$K.$ Indeed,
this call's payoff is equal to the amount of X 
\begin{equation*}
\left( \sqrt{f(T)}-\frac{K}{\sqrt{f(T)}}\right) _{+}
\end{equation*}%
which is equivalent to the amount of USD 
\begin{equation*}
\sqrt{f(T)}\left( \sqrt{f(T)}-\frac{K}{\sqrt{f(T)}}\right) _{+}=\left(
f(T)-K\right) _{+}
\end{equation*}%
as we can exchange $X$ for USD at the rate $\sqrt{f(T)}.$ Analogously, the
in-the-money payoff of $P_{\$/X}$ is equivalent to selling $\$1$ USD for
\euro $1/K$.

Further, by multiplying the price of $\mathcal{E}_{\text{\euro }/X}^{C}$
priced in X by $\sqrt{f(0)}$, we convert its option price in X to the price
in USD, and by multiplying the price of $\mathcal{E}_{\$/X}^{P}$ priced in X
by $1/\sqrt{f(0)},$ we convert its price to EUR. Hence 
\begin{eqnarray}
C_{\text{\euro }/\$}(0,T,f(0),K,r_{\$},r_{\text{\euro }}) &=&\sqrt{f(0)}%
\mathcal{E}_{\text{\euro }/X}^{C}(0,T,f(0),K,r_{\$},r_{\text{\euro }}),
\label{e1} \\
P_{\$/\text{\euro }}\left( 0,T,\frac{1}{f(0)},\frac{1}{K},r_{\text{\euro }%
},r_{\$}\right) &=&\frac{1}{\sqrt{f(0)}}\mathcal{E}_{\$/X}^{P}\left( 0,T,%
\frac{1}{f(0)},\frac{1}{K},r_{\text{\euro }},r_{\$}\right) .  \notag
\end{eqnarray}%
Comparing the resulting formulas for $C_{\text{\euro }/\$}$ and $P_{\$/\text{%
\euro }}$, it is not difficult to show that the foreign-domestic symmetry (%
\ref{eq:sym}) holds, which is also obvious from the duality principle (see 
\cite{NaF} and references therein).
\end{example}

Now let us consider a general FX option pricing formula based on the
intermediate currency. Let $S_{\text{\euro }/\$}(t)=f(t)$ be the EUR-USD
exchange rate at time $t$ defined on a filtered probability space $(\Omega ,{%
\mathcal{F}},\{{\mathcal{F}}_{t}\},Q^{X}),$ where $Q^{X}$ is an EMM
corresponding to the virtual market for which the intermediate currency X is
domestic (note that at the start of this subsection we demonstrated that
there is a broad class of models for which $Q^{X}$ exists). Assume that the
distribution of $f(t)$ is such that $f(t)$ and $1/f(t)$ have second moments.
We remark that we do not assume a particular model for $f(t)$ in the pricing
part of this section. For simplicity, let the interest rates for the USD and
EUR money markets, $r_{\$}$ and $r_{\text{\euro }}$, be constant. As we
mentioned earlier, the intermediate currency interest rate $r_{X}(t)$ is, in
general, not constant even when $r_{\$}$ and $r_{\text{\euro }}$ are
constant. We assume that $r_{X}(t)$ is adapted to the same filtration ${%
\mathcal{F}}_{t}$ and 
\begin{equation}
B_{X}(t)=\exp \left( \int_{0}^{t}r_{X}(s)ds\right) .  \label{BX}
\end{equation}%
Introduce the discounting factor $D_{X}(t,T)$ related to the intermediate
currency interest rate and the intermediate currency non-defaultable
zero-coupon bond price $P_{X}(t,T)$: 
\begin{equation}
D_{X}(t,T)=\exp \left( -\int_{t}^{T}r_{X}(s)ds\right)  \label{DX}
\end{equation}%
and 
\begin{equation}
P_{X}(t,T)=\mathbb{E}_{Q^{X}}\left[ D_{X}(t,T)|{\mathcal{F}}_{t}\right] ,
\label{PX}
\end{equation}%
where we assumed that $D_{X}(t,T)$ has finite moments. Since $Q^{X}$ is EMM,
the discounted $Y_{\text{\euro }/X}(t)$ and $Y_{\$/X}(t)$, 
\begin{equation*}
D_{X}(0,t)Y_{\text{\euro }/X}(t)=D_{X}(0,t)S_{\text{\euro }/X}(t)B_{\text{%
\euro }}(t)=D_{X}(0,t)\sqrt{f(t)}B_{\text{\euro }}(t)
\end{equation*}%
and 
\begin{equation*}
D_{X}(0,t)Y_{\$/X}(t)=D_{X}(0,t)S_{\$/X}(t)B_{\$}(t)=D_{X}(0,t)\frac{1}{%
\sqrt{f(t)}}B_{\$}(t),
\end{equation*}%
are $Q^{X}$-martingales. Hence we obtain for any $t\geq 0$ 
\begin{eqnarray*}
\sqrt{f(0)} &=&e^{r_{\text{\euro }}t}\mathbb{E}_{Q^{X}}\left[ D_{X}(0,t)%
\sqrt{f(t)}\right] , \\
\frac{1}{\sqrt{f(0)}} &=&e^{r_{\$}t}\mathbb{E}_{Q^{X}}\left[ \frac{D_{X}(0,t)%
}{\sqrt{f(t)}}\right] .
\end{eqnarray*}%
Therefore, to obey the no-arbitrage condition, the distribution of $f(t),$ $%
t\geq 0,$ under $Q^{X}$ should be so that 
\begin{equation}
\frac{\mathbb{E}_{Q^{X}}\left[ D_{X}(0,t)\sqrt{f(t)}\right] }{\mathbb{E}%
_{Q^{X}}\left[ \frac{D_{X}(0,t)}{\sqrt{f(t)}}\right] }=e^{(r_{\$}-r_{\text{%
\euro }})t}f(0).  \label{eq:arbit0}
\end{equation}

In option pricing we will consider the following natural class of payoff
functions $g(x;K),$ where $x>0$ denotes a price of the underlier and $K\geq
0 $ has the meaning of a strike.

\begin{assumption}
\label{def:g1}Let payoff functions $g(x;K)$ be homogeneous functions of
order $1,$ i.e. for any $a>0:$%
\begin{equation}
a\cdot g(x;K)=g(ax;aK).  \label{eq:g1}
\end{equation}
\end{assumption}

It is clear that e.g. plain vanilla puts and calls satisfy (\ref{eq:g1}).
For definiteness, assume that $g(x;K)$ is a payoff of an option written on
one EUR, where $x$ has the meaning of EUR-USD exchange rate, and $K$ and $g$
are denominated in USD. As in the case of a call (see Example~\ref{ex:32}),
the amount of USD $g(x;K)$ is equivalent to the amount $G(x;K)$ in X: 
\begin{equation*}
G(x;K):=\frac{1}{\sqrt{x}}g(x;K)=g\left( \sqrt{x};\frac{K}{\sqrt{x}}\right) ,
\end{equation*}%
where $1/\sqrt{x}$ has the meaning of the exchange rate USD-X (cf.
Definition~\ref{def:pseudocurrency}) and $G(x;K)$ and $K/\sqrt{x}$ are
denominated in $X$. According to the risk-neutral pricing theory, we can
write the value of the European option $V_{\text{\euro }/X}(t)$ with payoff $%
g(\sqrt{x};\frac{K}{\sqrt{x}})$ and maturity $T$ at time $t\leq T$ as 
\begin{equation*}
V_{\text{\euro }/X}(t)=\mathbb{E}_{Q^{X}}\left[ \left. D_{X}(t,T)g\left( 
\sqrt{f(T)};\frac{K}{\sqrt{f(T)}}\right) \right\vert {\mathcal{F}}_{t}\right]
.
\end{equation*}%
Note that this is an option on EUR priced in X. The price in dollars for
this option is 
\begin{equation}
V_{\text{\euro }/\$}(t)=\sqrt{f(t)}\mathbb{E}_{Q^{X}}\left[ \left.
D_{X}(t,T)g\left( \sqrt{f(T)};\frac{K}{\sqrt{f(T)}}\right) \right\vert {%
\mathcal{F}}_{t}\right] .  \label{eq:tQX1}
\end{equation}%
Analogously, we can derive a formula for an option on USD priced in EUR: 
\begin{equation}
V_{\$/\text{\euro }}(t)=\frac{1}{\sqrt{f(t)}}\mathbb{E}_{Q^{X}}\left[ \left.
D_{X}(t,T)g\left( \frac{1}{\sqrt{f(T)}};\sqrt{f(T)}K\right) \right\vert {%
\mathcal{F}}_{t}\right] ,  \label{eq:tQX2}
\end{equation}%
where $g(y;K)$ is a payoff of an option written on one USD, $y$ has the
meaning of USD-EUR exchange rate, and $K$ and $g$ are denominated in EUR. We
summarise this result in the following theorem.

\begin{theorem}
\label{thm:priceQX1}Assume that the EUR-USD exchange rate $f(t)$ satisfies a
model for which the no-arbitrage condition (\ref{eq:arbit0}) holds. Then the
arbitrage price of a European option on EUR with a payoff $g(x;K)$ being
homogeneous function of order $1$ and maturity time $T$ is given by (\ref%
{eq:tQX1}) and the arbitrage price of an option on USD is given by (\ref%
{eq:tQX2}).
\end{theorem}

It is not difficult to show that the foreign-domestic symmetry (\ref{eq:sym}%
) holds when we use the pricing formulas (\ref{eq:tQX1}) and (\ref{eq:tQX2})
based on the intermediate currency, which also follows from the duality
principle (see \cite{NaF} and references therein).

\subsection{T-forward measure for the intermediate market\label{sec:QXT}}

Introduce the T-forward measure $Q_{T}^{X}$ equivalent to $Q^{X}$ on ${%
\mathcal{F}}_{T}$ with the Radon-Nikodym derivative 
\begin{equation}
\frac{Q_{T}^{X}}{Q^{X}}=\frac{1}{P_{X}(0,T)B_{X}(T)}  \label{RN1}
\end{equation}%
and for $t>0$ 
\begin{equation}
E_{Q}\left[ \left. \frac{Q_{T}^{X}}{Q^{X}}\right\vert \mathcal{F}_{t}\right]
=\frac{P_{X}(t,T)}{P_{X}(0,T)B_{X}(t)}.  \label{RN2}
\end{equation}%
Under this forward measure, we get \cite{Jam89,Geman95} (see also \cite%
{Bjork})$:$ 
\begin{eqnarray}
\sqrt{f(0)} &=&e^{r_{\text{\euro }}T}\mathbb{E}_{Q^{X}}\left[ D_{X}(0,T)%
\sqrt{f(T)}\right] =e^{r_{\text{\euro }}T}P_{X}(0,T)\mathbb{E}_{Q_{T}^{X}}%
\left[ \sqrt{f(T)}\right] ,  \label{eq:new1} \\
\frac{1}{\sqrt{f(0)}} &=&e^{r_{\$}T}\mathbb{E}_{Q^{X}}\left[ \frac{D_{X}(0,T)%
}{\sqrt{f(T)}}\right] =e^{r_{\$}T}P_{X}(0,T)\mathbb{E}_{Q_{T}^{X}}\left[ 
\frac{1}{\sqrt{f(T)}}\right] .  \notag
\end{eqnarray}%
Then the no-arbitrage condition (\ref{eq:arbit0}) becomes 
\begin{equation}
\frac{\mathbb{E}_{Q_{T}^{X}}\left[ \sqrt{f(T)}\right] }{\mathbb{E}%
_{Q_{T}^{X}}\left[ \frac{1}{\sqrt{f(T)}}\right] }=e^{(r_{\$}-r_{\text{\euro }%
})T}f(0).  \label{eq:arbit}
\end{equation}%
Further, (\ref{eq:new1}) implies that the bond price $P_{X}(0,T)$ should
satisfy 
\begin{equation}
P_{X}(0,T)=e^{-r_{\text{\euro }}T}\frac{\sqrt{f(0)}}{\mathbb{E}_{Q_{T}^{X}}%
\sqrt{f(T)}}=e^{-r_{\$}T}\frac{1}{\sqrt{f(0)}\mathbb{E}_{Q_{T}^{X}}\left[ 
\frac{1}{\sqrt{f(T)}}\right] }\ .  \label{eq:rxx}
\end{equation}%
Note that $f(0)$ is the current EUR-USD exchange rate and hence it is
observable as well as $r_{\$}$ and $r_{\text{\euro }}$. The current forward
EUR-USD exchange rate 
\begin{equation}
F_{\text{\euro }/\$}(0,T)=e^{(r_{\$}-r_{\text{\euro }})T}f(0)
\label{eq:forward}
\end{equation}%
is also observable on the USD market.

We remark that the forward EUR-X and USD-X exchange rates, 
\begin{equation}
F_{\text{\euro }/X}(t,T)=e^{-r_{\text{\euro }}(T-t)}\frac{\sqrt{f(t)}}{%
P_{X}(t,T)}\text{ and }F_{\$/X}(t,T)=e^{-r_{\$}(T-t)}\frac{1}{P_{X}(t,T)%
\sqrt{f(t)}},  \label{eq:xxxforw}
\end{equation}%
are both $Q_{T}^{X}$-martingales. For convenience, we recall that if $%
r_{X}(t)$ is deterministic then the two measures $Q^{X}$ and $Q_{T}^{X}$
coincide.

It is also not difficult to show that 
\begin{eqnarray*}
\sqrt{f(t)} &=&e^{r_{\text{\euro }}(T-t)}\mathbb{E}_{Q^{X}}\left[ \left.
D_{X}(t,T)\sqrt{f(T)}\right\vert {\mathcal{F}}_{t}\right] =e^{r_{\text{\euro 
}}(T-t)}P_{X}(t,T)\mathbb{E}_{Q_{T}^{X}}\left[ \left. \sqrt{f(T)}\right\vert 
{\mathcal{F}}_{t}\right] , \\
\frac{1}{\sqrt{f(t)}} &=&e^{r_{\$}(T-t)}\mathbb{E}_{Q^{X}}\left[ \left. 
\frac{D_{X}(t,T)}{\sqrt{f(T)}}\right\vert {\mathcal{F}}_{t}\right] =e^{r_{%
\text{\$}}(T-t)}P_{X}(t,T)\mathbb{E}_{Q_{T}^{X}}\left[ \left. \frac{1}{\sqrt{%
f(T)}}\right\vert {\mathcal{F}}_{t}\right] .
\end{eqnarray*}%
Then 
\begin{equation}
P_{X}(t,T)=e^{-r_{\text{\euro }}(T-t)}\frac{\sqrt{f(t)}}{\mathbb{E}%
_{Q_{T}^{X}}\left[ \left. \sqrt{f(T)}\right\vert {\mathcal{F}}_{t}\right] }%
=e^{-r_{\$}(T-t)}\frac{1}{\sqrt{f(t)}\mathbb{E}_{Q_{T}^{X}}\left[ \left. 
\frac{1}{\sqrt{f(T)}}\right\vert {\mathcal{F}}_{t}\right] }\ .
\label{eq:rxx2}
\end{equation}

The pricing formula (\ref{eq:tQX1}) under the T-forward measure $Q_{T}^{X}$
becomes 
\begin{eqnarray}
V_{\text{\euro }/\$}(t) &=&\sqrt{f(t)}\mathbb{E}_{Q^{X}}\left[ \left.
D_{X}(t,T)g\left( \sqrt{f(T)};\frac{K}{\sqrt{f(T)}}\right) \right\vert {%
\mathcal{F}}_{t}\right]  \label{eq:QT1} \\
&=&\sqrt{f(t)}P_{X}(t,T)\mathbb{E}_{Q_{T}^{X}}\left[ \left. g\left( \sqrt{%
f(T)};\frac{K}{\sqrt{f(T)}}\right) \right\vert {\mathcal{F}}_{t}\right] 
\notag \\
&=&\frac{e^{-r_{\$}(T-t)}}{\mathbb{E}_{Q_{T}^{X}}\left[ \left. \frac{1}{%
\sqrt{f(T)}}\right\vert {\mathcal{F}}_{t}\right] }\mathbb{E}_{Q_{T}^{X}}%
\left[ \left. g\left( \sqrt{f(T)};\frac{K}{\sqrt{f(T)}}\right) \right\vert {%
\mathcal{F}}_{t}\right] ,  \notag
\end{eqnarray}%
where in the last line we used (\ref{eq:rxx2}). Analogously we have (see (%
\ref{eq:tQX2})): 
\begin{equation}
V_{\$/\text{\euro }}(t)=\frac{e^{-r_{\text{\euro }}(T-t)}}{\mathbb{E}%
_{Q_{T}^{X}}\left[ \left. \sqrt{f(T)}\right\vert {\mathcal{F}}_{t}\right] }%
\mathbb{E}_{Q_{T}^{X}}\left[ \left. g\left( \frac{1}{\sqrt{f(t)}};\sqrt{f(T)}%
K\right) \right\vert {\mathcal{F}}_{t}\right] .  \label{eq:QT2}
\end{equation}%
We summarize this result in the next theorem.

\begin{theorem}
\label{thm:main}Assume that the EUR-USD exchange rate $f(t)$ satisfies a
model for which the no-arbitrage condition (\ref{eq:arbit0}) or (\ref%
{eq:arbit}) holds. Then the arbitrage price of an option on EUR with a
payoff $g(x;K)$ being homogeneous function of order $1$and maturity time $T$
is given by (\ref{eq:QT1}) and the arbitrage price of an option on USD is
given by (\ref{eq:QT2}).
\end{theorem}

The benefit of (\ref{eq:QT1}) and (\ref{eq:QT2}) vs (\ref{eq:tQX1}) and (\ref%
{eq:tQX2}) is that in (\ref{eq:QT1}) and (\ref{eq:QT2}) we do not need to
compute the intermediate currency interest rate $r_{X}(t).$

\begin{example}
\label{ex:33}The prices of the call for buying \euro $1$ for $\$K$ and of
the put for selling $\$1$ for \euro $1/K$ are equal to 
\begin{align}
C_{\text{\euro }/\$}(0,T,K)& =\frac{e^{-r_{\$}T}}{\mathbb{E}_{Q_{T}^{X}}%
\left[ \frac{1}{\sqrt{f(T)}}\right] }\mathbb{E}_{Q_{T}^{X}}\left[ \left( 
\sqrt{f(T)}-\frac{K}{\sqrt{f(T)}}\right) _{+}\right] ,  \label{eq:thm1} \\
P_{\$/\text{\euro }}\left( 0,T,\frac{1}{K}\right) & =\frac{e^{-r_{\text{%
\euro }}T}}{\mathbb{E}_{Q_{T}^{X}}\sqrt{f(T)}}\mathbb{E}_{Q_{T}^{X}}\left[
\left( \frac{\sqrt{f(T)}}{K}-\frac{1}{\sqrt{f(T)}}\right) _{+}\right] . 
\notag
\end{align}
\end{example}

To conclude, we derived the consistent pricing formulas for FX options.
Although the formulas are derived using the virtual X market, their
evaluation depends on parameters of the USD and EUR markets only. When we
are interested in option prices at the current time $t=0,$ they are valid
for any distribution (i.e., we do not need to explicitly define the process $%
f(t))$ of the exchange rate $f(T)$ which satisfies (\ref{eq:arbit}). We will
demonstrate this observation in illustrations of these pricing formulas in
Section~\ref{sec:ill}.

\section{Extension to the multi--currencies case\label{sec:multiN}}

In this section we extend the approach of pricing FX options developed in
the previous section to the multi-currencies case. Let us assume we have $N$
currencies $c_{i}$, where $i=1,\dots ,N$. Fixing one currency, for
definiteness $i=N$, we can introduce the $N-1$ exchange rates 
\begin{equation}
f_{j}=S_{c_{j}/c_{N}}>0,\quad j=1,\ldots ,N-1,  \label{eq:exchangerateMulti}
\end{equation}%
which denote the exchange rates between the currency $c_{N}$ to all other
currencies $c_{i},$ $i=1,\dots ,N-1$.

Now we introduce the intermediate currency $X$ by defining the $N$ exchange
rates $S_{c_{i}/X}$ as follows 
\begin{equation}
S_{c_{i}/X}=f_{1}^{b_{i1}}\times f_{2}^{b_{i2}}\times \cdots \times
f_{N-1}^{b_{iN-1}},\quad i=1,\ldots ,N,  \label{eq:intermediateExchangeMulti}
\end{equation}%
where $b_{ij}\in \mathbb{R}$ are so that 
\begin{eqnarray*}
b_{ii} &=&1-\alpha _{i},\ i=1,\ldots ,N, \\
b_{ij} &=&-\alpha _{j},\ i\neq j,\ \ i,j=1,\ldots ,N.
\end{eqnarray*}%
By symmetry arguments (see also Remark~\ref{rem:multi}\ below), we choose 
\begin{equation}
\alpha _{i}=\frac{1}{N},\ \ i=1,\ldots ,N-1.  \label{alpha}
\end{equation}%
Note that $S_{c_{i}/X}$ is the exchange rate between the observable currency 
$c_{i}$ and the introduced intermediate currency $X$ and hence it is the
worth of $1$ unit of currency $c_{i}$ in the intermediate currency $X$. In
the case (\ref{eq:exchangerateMulti})-(\ref{eq:intermediateExchangeMulti}), (%
\ref{alpha}), the exchange rate $S_{c_{i}/X}$ can be written in the concise
form via geometric mean $GM(f_{j})$ of the sequence of $f_{j}:$ 
\begin{equation*}
S_{c_{i}/X}=f_{i}\left[ \left( \prod_{j=1}^{N-1}\frac{1}{f_{j}}\right)
^{1/(N-1)}\right] ^{(N-1)/N}:=f_{i}\left[ GM(f_{j})\right] ^{(N-1)/N}.
\end{equation*}

We assume that the currency market under a `market' measure $P$ is described
by the system: 
\begin{equation}
\begin{aligned} df_j &= \mu_j(t) f_j dt + \sigma_j(t) f_j d\tilde{W}_j,\quad
j = 1,\ldots,N-1, \\ d\tilde{W}_l d\tilde{W}_k &=d\tilde{W}_k d\tilde{W}_l =
\rho_{lk}(t)dt,\quad l,k = 1,\ldots,N-1, \end{aligned}
\label{eq:SDEsystemMulti}
\end{equation}%
and 
\begin{equation}
\begin{aligned} dB_{i} &= r_{i}(t)B_{i}dt,\quad i = 1,\ldots,N,\\
\end{aligned}  \label{eq:Bi}
\end{equation}%
where $B_{i}(t)$ describes the bank account of currency $c_{i}$ with its
short rate $r_{i}(t)$; $\sigma _{j}(t)>0$ is the volatility of the exchange
rate $f_{j}(t)$, $\mu _{j}(t)$ is its drift; and $\tilde{W}(t)=(\tilde{W}%
_{1}(t),\ldots ,\tilde{W}_{N-1}(t))^{T}$ is an $N-1$--dimensional correlated
Wiener process with the correlation matrix $R(t)\in \mathbb{R}^{N-1\times
N-1}$ which components we denote by $\rho _{ij}(t)$ (obviously $\rho
_{ii}=1).$ It is assumed that $r_{i}(t)$, $\sigma _{j}(t)$, $\mu _{j}(t)$
are stochastic processes adapted to a filtration $\mathcal{F}_{t}$ to which $%
\tilde{W}(t)$ is also adapted, and they have bounded second moments and $%
\sigma _{j}(t)$ satisfy Novikov's condition. Furthermore, let us assume that
the matrix $R$ is symmetric strictly positive definite. Then using the
Cholesky decomposition, we can represent $R=LL^{T}$, where $L\in \mathbb{R}%
^{N-1\times N-1}$ is a lower triangular matrix with entries $L_{i,j}$. Using
this decomposition, we can rewrite the SDEs (\ref{eq:SDEsystemMulti}) as 
\begin{equation}
df_{j}=\mu _{j}(t)f_{j}dt+\sigma
_{j}(t)f_{j}\sum\limits_{k=1}^{j}L_{jk}(t)dW_{k},\quad j=1,\ldots ,N-1,
\label{eq:model11}
\end{equation}%
where 
\begin{equation*}
L_{ii}(t)=\sqrt{1-\sum\limits_{k=1}^{i-1}L_{ik}^{2}(t)},\quad L_{ji}(t)=%
\frac{\rho _{ij}-\sum\limits_{k=1}^{i-1}L_{jk}(t)L_{ik}(t)}{L_{ii}(t)},\quad 
\text{for }j>i,
\end{equation*}%
and $W(t)=(W_{1}(t),\ldots ,W_{N-1}(t))^{T}$ is an $N-1$--dimensional
standard Wiener process. We first show that the intermediate currency
introduced in (\ref{eq:intermediateExchangeMulti}) permits an arbitrage-free
market involving all $N$ currencies.

\begin{theorem}
\label{thm:multicurrencyIntermediate} Assume that $N-1$ exchange rates $%
f_{j} $ between the currency $c_{N}$ to all other currencies $c_{i},$ $%
i=1,\dots ,N-1,$ under a `market' measure are described by the model (\ref%
{eq:model11}) together with (\ref{eq:Bi}). Consider the intermediate
currency $X$ introduced in (\ref{eq:intermediateExchangeMulti}). There is
the unique intermediate currency interest rate $r_{X}(t)$ defined by 
\begin{equation}
r_{X}(t)=\frac{1}{N}\sum\limits_{i=1}^{N}r_{i}(t)+\frac{1}{2N}\left( 1-\frac{%
1}{N}\right) \sum\limits_{i=1}^{N-1}\sigma _{i}^{2}(t)-\frac{1}{N^{2}}%
\sum\limits_{k=1}^{N-1}\sum\limits_{j=1}^{k-1}\sigma _{j}(t)\sigma
_{k}(t)\rho _{jk}(t)  \label{eq:rxxxm}
\end{equation}%
and there is an EMM $Q^{X}$ for the intermediate pseudo-currency market,
i.e., under (\ref{eq:rxxxm}) this market is arbitrage-free.
\end{theorem}

\noindent \textbf{Proof}. Applying the Ito formula to (\ref%
{eq:intermediateExchangeMulti}), we obtain the SDEs for the exchange rates $%
S_{c_{i}/X}$: 
\begin{align*}
\frac{dS_{c_{i}/X}}{S_{c_{i}/X}}& =\left[ \frac{1}{N}\sum\limits_{j=1}^{N-1}%
\left( \frac{1}{2}\left( \frac{1}{N}+1\right) \sigma _{j}^{2}-\sigma _{i}%
\mathbbm{1}_{i\neq N}\sigma _{j}\rho _{ij}+\frac{1}{N}\sigma
_{j}\sum\limits_{k=1}^{j-1}\sigma _{k}\rho _{kj}-\mu _{j}\right) +\mu _{i}%
\mathbbm{1}_{i\neq N}\right] dt \\
& \quad -\frac{1}{N}\sum\limits_{j=1}^{N-1}\sum\limits_{k=1}^{j}\sigma
_{j}L_{jk}dW_{k}+\sigma _{i}\mathbbm{1}_{i\neq
N}\sum\limits_{k=1}^{i}L_{ij}dW_{k},\quad i=1,\ldots ,N.
\end{align*}%
On the considered market the risky assets have the prices $%
Y_{c_{i}/X}=S_{c_{i}/X}B_{i},$ $i=1,\ldots ,N$. Introduce the discounted
risky assets' prices in the usual way: 
\begin{equation}
\tilde{Y}_{c_{i}/X}(t)=\frac{S_{c_{i}/X}(t)B_{i}(t)}{B_{X}(t)},\quad
i=1,\ldots ,N.  \label{tiY}
\end{equation}%
The discounted prices satisfy the SDEs%
\begin{align*}
\frac{d\tilde{Y}_{c_{i}/X}}{\tilde{Y}_{c_{i}/X}}& =\left[ r_{i}-r_{X}\right]
dt \\
& +\left[ \frac{1}{N}\sum\limits_{j=1}^{N-1}\left( \frac{1}{2}\left( \frac{1%
}{N}+1\right) \sigma _{j}^{2}-\sigma _{i}\mathbbm{1}_{i\neq N}\sigma
_{j}\rho _{ij}+\frac{1}{N}\sigma _{j}\sum\limits_{k=1}^{j-1}\sigma _{k}\rho
_{kj}-\mu _{j}\right) +\mu _{i}\mathbbm{1}_{i\neq N}\right] dt \\
& \quad -\frac{1}{N}\sum\limits_{j=1}^{N-1}\sum\limits_{k=1}^{j}\sigma
_{j}L_{jk}dW_{k}+\sigma _{i}\mathbbm{1}_{i\neq
N}\sum\limits_{k=1}^{i}L_{ik}dW_{k},\quad i=1,\ldots ,N.
\end{align*}%
The no-arbitrage condition requires existence of an EMM $Q^{X}$ under which
all $\tilde{Y}_{c_{i}/X}$ are martingales. This implies that for $Q^{X}$ to
exist the following system of $N$ simultaneous linear algebraic equations in 
$N$ unknown variables (which are the market prices of risk $\gamma _{k}$, $%
k=1,\ldots ,N-1$, and $r_{X})$ should have a solution: 
\begin{eqnarray}
&&r_{i}-r_{X}+\frac{1}{N}\sum\limits_{j=1}^{N-1}\left( \frac{1}{2}\left( 
\frac{1}{N}+1\right) \sigma _{j}^{2}-\sigma _{i}\mathbbm{1}_{i\neq N}\sigma
_{j}\rho _{ij}+\frac{1}{N}\sigma _{j}\sum\limits_{k=1}^{j-1}\sigma _{k}\rho
_{kj}-\mu _{j}\right) +\mu _{i}\mathbbm{1}_{i\neq N}\ \ \ \ \ \ \ \ \ \ \ \ 
\label{eq:SEQmult} \\
&=&-\frac{1}{N}\sum\limits_{j=1}^{N-1}\sum\limits_{k=1}^{j}\sigma
_{j}L_{jk}\gamma _{k}+\sigma _{i}\mathbbm{1}_{i\neq
N}\sum\limits_{k=1}^{i}L_{ik}\gamma _{k},\quad i=1,\ldots ,N.  \notag
\end{eqnarray}%
Subtracting the equation (\ref{eq:SEQmult}) with $i=N$ from the equations (%
\ref{eq:SEQmult}) for $i\neq N,$ we obtain%
\begin{equation}
r_{i}-r_{N}+\mu _{i}-\frac{1}{N}\sigma _{i}\sum\limits_{k=1}^{N-1}\sigma
_{k}\rho _{ik}=\sigma _{i}\sum\limits_{k=1}^{i}L_{ik}\gamma _{k},\quad
i=1,\ldots ,N-1.  \label{aga1}
\end{equation}%
Using (\ref{aga1}), we recurrently find the market prices of risk:%
\begin{equation}
\gamma _{i}=\frac{r_{i}-r_{N}+\mu _{i}-\frac{1}{N}\sigma _{i}^{2}-\frac{1}{N}%
\sigma _{i}\sum\limits_{k=1}^{N-1}\sigma _{k}\rho _{ik}-\sigma
_{i}\sum\limits_{k=1}^{i-1}L_{i,k}\gamma _{k}}{\sigma _{i}L_{i,i}},\quad
i=1,\ldots ,N-1,  \label{aga2}
\end{equation}%
which are well defined because due to our assumptions $\sigma _{i}>0$ and $%
L_{i,i}>0.$ Further, sum up (\ref{aga1}) over $i$ from $i=1$ to $N-1$ and
substitute the result in (\ref{eq:SEQmult}) with $i=N$ to confirm (\ref%
{eq:rxxxm}). The found $\gamma _{i},$ $i=1,\ldots ,N-1,$ from (\ref{aga2})
and $r_{X}$ from (\ref{eq:rxxxm}) together with Girsanov's theorem ensure
that there is an EMM $Q^{X}$ under which all $\tilde{Y}_{c_{i}/X}$ are
martingales. Hence, the considered market is arbitrage free. Theorem~\ref%
{thm:multicurrencyIntermediate} is proved. $\square $

\begin{remark}
\label{rem:multi} Recall that we chose to use $\alpha _{1}=\cdots =\alpha
_{N-1}=\frac{1}{N}$ in (\ref{eq:intermediateExchangeMulti}). If we repeat
the proof of Theorem \ref{thm:multicurrencyIntermediate} for arbitrary $%
0<\alpha _{j}<1$ then we arrive at the following intermediate currency
interest rate $r_{X}:$ 
\begin{equation}
r_{X}=\left( 1-\sum\limits_{j=1}^{N-1}\alpha _{j}\right)
r_{N}+\sum\limits_{j=1}^{N-1}\alpha _{j}r_{j}+\sum\limits_{j=1}^{N-1}\frac{%
\alpha _{j}(1-\alpha _{j})}{2}\sigma
_{j}^{2}-\sum\limits_{j=1}^{N-1}\sum\limits_{k=1}^{j-1}\sigma _{j}\alpha
_{j}\alpha _{k}\sigma _{k}\rho _{jk}.  \label{aga3}
\end{equation}%
ensuring that there is an EMM in this market. We see that the choice $\alpha
_{j}=\frac{1}{N}$ results in the symmetry so that each $r_{j}$ enters (\ref%
{aga3}) with the same weight. Other choices of $\alpha _{j}$ give a
`preference' to a particular currency(ies) and can be exploited for
calibration purposes but it is not considered here.
\end{remark}

Analogously to Assumption~\ref{def:g1}, we will consider payoffs as
first-order homogeneous functions the multi-currencies case.

\begin{assumption}
\label{def:g2}Let payoff functions $g(x_{1},\ldots ,x_{N-1};K)$ be
homogeneous functions of order $1,$ i.e. for any $a>0$%
\begin{equation}
a\cdot g(x_{1},\ldots ,x_{N-1};K)=g(ax_{1},\ldots ,ax_{N-1};aK).
\label{aga4}
\end{equation}
\end{assumption}

Most of multi-currency options (e.g. basket options \cite{Clark}) have
payoffs belonging to this class. Consider a European-type option with
maturity time $T$ and payoff in the currency $c_{N}$: 
\begin{equation*}
g(T):=g(f_{1}(T),\ldots ,f_{N-1}(T);K).
\end{equation*}%
Its equivalent value in the intermediate currency $X$ is equal to (see (\ref%
{eq:intermediateExchangeMulti})):%
\begin{align}
G(T):=& S_{c_{N}/X}(T)\cdot g(T)=g\Big(f_{1}(T)S_{c_{N}/X}(T),\ldots
,f_{N-1}(T)S_{c_{N}/X}(T);K\cdot S_{c_{N}/X}(T)\Big)  \label{aga5} \\
=& g\Big(S_{c_{1}/X}(T),\ldots ,S_{c_{N-1}/X}(T);K\cdot S_{c_{N}/X}(T)\Big)=g%
\Big(S_{c_{1}/X}(T),\ldots ,S_{c_{N}/X}(T);K^{\prime }\Big),  \notag
\end{align}%
where $K^{\prime }=K\cdot S_{c_{N}/X}(t)$ is the equivalent strike in $X$.
It is not difficult to see that at the maturity time $T$ the option holder
is indifferent between receiving $g(T)$ in currency $c_{N}$ or $G(T)$ in
currency $X$ as he can obtain the same amount by exchanging $G(T)$ to $c_{N}$%
: 
\begin{eqnarray*}
\frac{G(T)}{S_{c_{N}/X}(T)} &=&\frac{1}{S_{c_{N}/X}(T)}g\Big(%
S_{c_{1}/X}(T),\ldots ,S_{c_{N-1}/X}(T);K\cdot S_{c_{N}/X}(T)\Big) \\
&=&g\Big(f_{1}(T),\ldots ,f_{N-1}(T);K\Big).
\end{eqnarray*}

\begin{example}[Basket option]
\label{ex:basket} Consider a basket option on the $c_{N}$ market written on
all $N-1$ exchange rates $f_{i}(t),$ $i=1,\ldots ,N-1,$ which pay-off
function is of the form \cite{Clark}: 
\begin{equation*}
g(x_{1},\ldots ,x_{N-1};K)=\left( \sum\limits_{i=1}^{N-1}\omega
_{i}x_{i}-K\right) _{+},
\end{equation*}%
where $x_{i},$ $i=1,\ldots ,N-1,$ and $K$ are denominated in the currency $%
c_{N}$ and $\omega _{i}\geq 0,$ $i=1,\ldots ,N-1,$ are some weights. The
equivalent pay-off on the $X$ currency market at the maturity $T$ is equal
to 
\begin{align}
G(T)& =S_{c_{N}/X}(T)\cdot g(f_{1}(T),\ldots ,f_{N-1}(T);K)
\label{eq:intermediatePayoff} \\
& =S_{c_{N}/X}(T)\left( \sum\limits_{i=1}^{N-1}\omega _{i}f_{i}(T)-K\right)
_{+}=\left( \sum\limits_{i=1}^{N-1}\omega _{i}S_{c_{i}/X}(t)-K\cdot
S_{c_{N}/X}(t)\right) _{+}  \notag \\
& =\left( \sum\limits_{i=1}^{N-1}\omega _{i}S_{c_{i}/X}(t)-K^{\prime
}\right) _{+},  \notag
\end{align}%
where $S_{c_{i}/X}(t)$ and $K^{\prime }$ are denominated in the intermediate
currency $X$.
\end{example}

As in the case of a single FX pair (see Theorem~\ref{prop:QX}), we have
demonstrated by Theorem~\ref{thm:multicurrencyIntermediate} that there is a
sufficiently broad class of models for which there is an EMM $Q^{X}$ with an
appropriate choice of the intermediate currency interest rate $r_{X}(t).$ We
now generalize the pricing formulas of Theorems~\ref{thm:priceQX1}\ and~\ref%
{thm:main} from a single FX pair to the multi-currency case.

Let the exchange rates $f_{i}(t)$ between the currency $c_{N}$ to all other
currencies $c_{i},$ $i=1,\ldots ,N-1,$ be defined on a filtered probability
space $(\Omega ,{\mathcal{F}},\{{\mathcal{F}}_{t}\},Q^{X}),$ where $Q^{X}$
is an EMM corresponding to the virtual market for which the intermediate
currency X is domestic. Assume that $f_{i}(t),$ $i=1,\ldots ,N-1,$ and the
exchange rates $S_{c_{i}/X}$ between the pseudo-currency $X$ to all the
currencies $c_{i},$ $i=1,\ldots ,N,$ defined in (\ref%
{eq:intermediateExchangeMulti}), (\ref{alpha}) have second moments. Also,
assume that $r_{X}(t)$ is adapted to the same filtration ${\mathcal{F}}_{t}$
and recall the expressions and assumptions for the money market account $%
B_{X}(t)$ (see (\ref{BX})), the discounting factor $D_{X}(t,T)$ related to
the intermediate currency interest rate (see (\ref{DX})) and the
intermediate currency zero-coupon bond price $P_{X}(t,T)$ (see (\ref{PX})).

Since $Q^{X}$ is an EMM, the discounted $Y_{c_{i}/X}(t)$ for all $i=1,\ldots
,N$, 
\begin{equation*}
\tilde{Y}%
_{c_{i}/X}=D_{X}(0,t)Y_{c_{i}/X}(t)=D_{X}(0,t)S_{c_{i}/X}(t)B_{c_{i}}(t),%
\quad i=1,\ldots ,N,
\end{equation*}%
are $Q^{X}$-martingales. Hence we obtain 
\begin{equation*}
S_{c_{i}/X}(0)=e^{r_{i}t}\mathbb{E}_{Q^{X}}\left[ D_{X}(0,t)S_{c_{i}/X}(t)%
\right] ,\quad i=1,\ldots ,N.
\end{equation*}%
Therefore, for all $i=1,\ldots ,N-1$ and $t>0,$ we have 
\begin{equation}
\frac{\mathbb{E}_{Q^{X}}\left[ D_{X}(0,t)S_{c_{i}/X}(t)\right] }{\mathbb{E}%
_{Q^{X}}\left[ D_{X}(0,t)S_{c_{N}/X}(t)\right] }=e^{(r_{i}-r_{N})t}\frac{%
S_{c_{i}/X}(0)}{S_{c_{N}/X}(0)}=e^{(r_{N}-r_{i})t}f_{i}(0).
\label{eq:arbit0multi}
\end{equation}%
Thus, to obey the no-arbitrage condition, the distributions of $%
S_{c_{i}/X}(t)$, $t>0,$ under $Q^{X}$ should be so that (\ref{eq:arbit0multi}%
) holds.

Consider a European option with maturity $T$ and pay-off function $G(T)$ on
the intermediate currency market. Its price in $X$ is equal to%
\begin{equation}
V_{X}(t)=\mathbb{E}_{Q^{X}}\left[ D_{X}(0,T)G(T)|{\mathcal{F}}_{t}\right] .
\label{eq:OptionpricingQX}
\end{equation}%
Using (\ref{aga5}), we obtain the price for this option in the currency $%
c_{N}$: 
\begin{equation}
V_{c_{N}}(t)=\frac{1}{S_{c_{N}/X}(t)}\mathbb{E}_{Q^{X}}\left[
D_{X}(0,T)\cdot g(S_{c_{1}/X}(T),\ldots ,S_{c_{N-1}/X}(T);KS_{c_{N}/X}(T))|{%
\mathcal{F}}_{t}\right] .  \label{aga6}
\end{equation}%
Then the analog of Theorem\ \ref{thm:priceQX1} is as follows.

\begin{theorem}
\label{thm:priceQX1m}Assume that the exchange rates $f_{i}(t),$ $i=1,\ldots
,N-1,$ (or $S_{c_{i}/X}(t),$ $i=1,\ldots ,N)$ satisfy a model for which the
no-arbitrage condition (\ref{eq:arbit0multi}) holds. Then on the $c_{N}$
market the arbitrage price $V_{c_{N}}(t)$ of a European option on $c_{1},$
\ldots , $c_{N-1}$ currencies with a first-order homogeneous payoff function 
$g(x_{1},\ldots ,x_{N-1};K)$ and maturity time $T$ is given by (\ref{aga6}).
\end{theorem}

Introduce the T-forward measure $Q_{T}^{X}$ equivalent to $Q^{X}$ on ${%
\mathcal{F}}_{T}$ with the Radon-Nikodym derivative as in (\ref{RN1}) (see
also (\ref{RN2})). Under this forward measure, we get 
\begin{equation}
S_{c_{i}/X}(0)=e^{r_{i}T}\mathbb{E}_{Q^{X}}\left[ D_{X}(0,T)S_{c_{i}/X}(T)%
\right] =e^{r_{i}T}P_{X}(0,T)\mathbb{E}_{Q_{T}^{X}}\left[ S_{c_{i}/X}(T)%
\right] ,\quad i=1,\ldots ,N.  \label{aga7}
\end{equation}%
Then the no-arbitrage conditions (\ref{eq:arbit0multi}) become 
\begin{equation}
\frac{\mathbb{E}_{Q_{T}^{X}}\left[ S_{c_{i}/X}(T)\right] }{\mathbb{E}%
_{Q_{T}^{X}}\left[ S_{c_{N}/X}(T)\right] }=e^{(r_{N}-r_{i})T}f_{i}(0),\quad
i=1,\ldots ,N-1.  \label{eq:arbitmulti}
\end{equation}%
Here $F_{c_{i}/c_{N}}(0)=e^{(r_{N}-r_{i})T}f_{i}(0)$ is the current forward $%
c_{i}$-$c_{N}$ exchange rate. It follows from (\ref{eq:arbitmulti}) that for
any $j=1,\ldots ,N:$%
\begin{equation}
\frac{\mathbb{E}_{Q_{T}^{X}}\left[ S_{c_{i}/X}(T)\right] }{\mathbb{E}%
_{Q_{T}^{X}}\left[ S_{c_{j}/X}(T)\right] }=e^{(r_{N}-r_{i})T}f_{i}(0),\quad
i=1,\ldots ,N,\ i\neq j.  \label{eq:arbitmulti2}
\end{equation}%
We remark that the no-arbitrage condition does not depend on the choice of $%
c_{N}$ used in (\ref{eq:exchangerateMulti}).

Further, (\ref{aga7}) implies that the bond price $P_{X}(0,T)$ should
satisfy 
\begin{equation}
P_{X}(0,T)=e^{-r_{i}T}\frac{S_{c_{i}/X}(0)}{\mathbb{E}%
_{Q_{T}^{X}}S_{c_{i}/X}(T)},\quad i=1,\ldots ,N.  \label{aga8}
\end{equation}%
We observe that the relationships (\ref{eq:arbitmulti}) ensure that (\ref%
{aga8}) holds for all $i=1,\ldots ,N.$ Note that $f_{i}(0)$ are the current $%
c_{i}/c_{N}$ exchange rates and hence $S_{c_{i}/X}(0)$ (see (\ref%
{eq:intermediateExchangeMulti})) are observable as well as all $r_{i}$.
Similarly to (\ref{eq:rxx2}), we also have 
\begin{equation*}
P_{X}(t,T)=e^{-r_{i}(T-t)}\frac{S_{c_{i}/X}(t)}{\mathbb{E}_{Q_{T}^{X}}\left[
S_{c_{i}/X}(T)|{\mathcal{F}}_{t}\right] },\quad i=1,\ldots ,N.
\end{equation*}

Analogously to (\ref{eq:QT2}), the pricing formula (\ref{aga6}) under the
T-forward measure $Q_{T}^{X}$ becomes 
\begin{equation}
V_{c_{N}}(t)=\frac{e^{-r_{N}(T-t)}}{\mathbb{E}_{Q_{T}^{X}}\left[
S_{c_{N}/X}(T)|{\mathcal{F}}_{t}\right] }\mathbb{E}_{Q_{T}^{X}}\left[
g(S_{c_{1}/X}(T),\ldots ,S_{c_{N-1}/X}(T);KS_{c_{N}/X}(T))|{\mathcal{F}}_{t}%
\right] .  \label{eq:thm1Multi2}
\end{equation}%
Then the analog of Theorem\ \ref{thm:main} is as follows.

\begin{theorem}
\label{thm:multicurrency}Assume that the exchange rates $f_{i}(t),$ $%
i=1,\ldots ,N-1,$ (or $S_{c_{i}/X}(t),$ $i=1,\ldots ,N)$ satisfy a model for
which the no-arbitrage condition (\ref{eq:arbitmulti}) (or (\ref%
{eq:arbit0multi})) holds. Then on the $c_{N}$ market the arbitrage price $%
V_{c_{N}}(t)$ of a European option on $c_{1},$ \ldots , $c_{N-1}$ currencies
with a first-order homogeneous payoff function $g(x_{1},\ldots ,x_{N-1};K)$
and maturity time $T$ is given by (\ref{eq:thm1Multi2}).
\end{theorem}

It is clear that the pricing formula (\ref{eq:thm1Multi2}) remains true if
we replace the currency $c_{N}$ with any other $c_{j}$ and the first-order
homogeneous payoff function $g(x_{1},\ldots ,x_{j-1},x_{j+1},\ldots ,x_{N};K)
$ is denominated in $c_{j}.$ We now return to Example~\ref{ex:basket}.

\begin{example}[Basket option pricing]
\label{ex:basketpricing} Let us make the same assumptions as in Example~\ref%
{ex:basket} and find the arbitrage price of a European option with pay-off $%
G(T)$ at maturity $T$ on the $X$ currency market given by 
\begin{equation*}
G(t)=\left( \sum\limits_{i=1}^{N-1}\omega _{i}S_{c_{i}/X}(t)-K\right) _{+},
\end{equation*}%
where $S_{c_{i}/X}(t)$ and $K$ are denominated in $X$. Following Theorem~\ref%
{thm:multicurrency}, the price of this basket option at time $0$,
denominated in currency $c_{N}$, is equal to 
\begin{equation*}
BasketOption_{c_{N}}(0)=\frac{e^{-r_{N}T}}{\mathbb{E}%
_{Q_{T}^{X}}S_{c_{N}/X}(T)}\mathbb{E}_{Q_{T}^{X}}\left[ \left(
\sum\limits_{i=1}^{N-1}\omega _{i}S_{c_{i}/X}(T)-K\right) _{+}\right] .
\end{equation*}
\end{example}

To conclude, we derived consistent pricing formulas (\ref{aga6}) and (\ref%
{eq:thm1Multi2}) for FX options in the multi-currency case. As it was in
Section~\ref{sec:intpricing}\ for a single FX pair, here in the
multi-currency case, although the pricing formulas~(\ref{aga6}) and~(\ref%
{eq:thm1Multi2}) are derived using the virtual X market, their evaluation
depends on parameters of the real $c_{i},$ $i=1,\ldots ,N,$ markets only.
The distinguishing feature of the considered approach is that we can price
all FX options regardless from their domestic market using the same measure
which in turn guarantees that all natural relationships between exchange
rates and FX options are automatically fulfilled. Consequently, using $%
N(N-1)/2$ smiles out of $N(N-1)$ available, we can calibrate a
multi-currency model to be used \textit{with the same parameters} for
pricing options on all the markets, which is convenient and desirable in
practice. 

\section{Illustrations\label{sec:ill}}

For illustrative purposes, we consider four examples in this section. The
first example (Section~\ref{sec:hes}) illustrates the use of FX pricing from
Section~\ref{sec:intpricing} in the case when the EUR-USD exchange rate $%
f(t) $ is described by the Heston model \cite{Hes} while the second example
(Section~\ref{sec:sabr}) deals with the SABR model \cite{hagan2002managing}.
In these two examples we follow the traditional route: we start with models
written under a `market' measure, then find an EMM $Q^{X}$ on the
intermediate currency market and use Theorem~\ref{thm:priceQX1}\ for pricing
FX options. The third example presented in Section~\ref{sec:esn} follows a
different route: we suggest a distribution for an exchange rate at maturity
time $T,$ e.g. for EUR-USD, under a forward measure $Q_{T}^{X}$ on the
intermediate currency market so that the no-arbitrage condition (\ref%
{eq:arbit}) is satisfied. Then we use Theorem~\ref{thm:main} or Theorem~\ref%
{thm:multicurrency} for pricing FX options. To this end, in Section~\ref%
{sec:esn} we assume that the EUR-USD exchange rate $f(T)$ has a skew normal
distribution. We remark that the use of the considered extended skew normal
model for FX pricing is novel. In Section~\ref{sec:mf}, we illustrate the
results of Sections~\ref{sec:intpricing} and ~\ref{sec:multiN} in the case
of the model-free approach \cite{Aus11,Aus14}.

\subsection{Heston model\label{sec:hes}}

For simplicity, let the interest rates for the USD and EUR money markets, $%
r_{\$}$ and $r_{\text{\euro }}$, be constant. Consider the Heston stochastic
volatility model for the EUR-USD exchange rate $S_{\text{\euro }/\$}(t)=f(t)$
written under a `market' measure\ \cite{Hes}: 
\begin{eqnarray}
df &=&\mu f\,dt+\,\sqrt{v}f\left( \sqrt{1-\rho ^{2}}\,dW_{1}(t)+\,\rho
dW_{2}(t)\right) ,\ \ f(0)=f_{0},  \label{hes1} \\
dv &=&\kappa \,(\theta -v)dt+\delta \sqrt{v}dW_{2}(t),\ v(0)=v_{0},  \notag
\end{eqnarray}%
where $W_{1}(t)$ and $W_{2}(t)$ are independent standard Wiener processes; $%
\sigma (t)=\sqrt{v(t)}\,$ is a (stochastic) volatility; $\theta ,$ $\kappa ,$
$\delta ,$ $f_{0}$ and $v_{0}$ are positive constants, satisfying 
\begin{equation}
2\kappa \theta \geq \delta ^{2};  \label{hes2}
\end{equation}%
and the correlation coefficient $\rho \in (-1,1).$ Recall that the condition
(\ref{hes2}) guarantees that zero is unattainable by $v(t)$ in finite time.

Following Section~\ref{sec:QX}, to re-write (\ref{hes1}) under $Q^{X},$ we
need to find the market prices of risk, $\gamma _{1}(t)$ and $\gamma _{2}(t)$%
, so that (cf. (\ref{eq:gamX})): 
\begin{equation}
\sqrt{1-\rho ^{2}}\gamma _{1}(t)+\rho \gamma _{2}(t)=\frac{\mu -v(t)/2+r_{%
\text{\euro }}-r_{\$}}{\sqrt{v(t)}}.  \label{hes3}
\end{equation}%
As it is standard for the Heston model \cite{Hes}, to deal with
incompleteness of the market, we choose 
\begin{equation}
\gamma _{2}(t)=\lambda \sqrt{v(t)},  \label{hes4}
\end{equation}%
where $\lambda $ is a constant. Therefore, we have 
\begin{eqnarray}
d\sqrt{f} &=&\left( r_{X}(t)-r_{\text{\euro }}\right) \sqrt{f}dt+\frac{\sqrt{%
v}}{2}\sqrt{f}\left( \sqrt{1-\rho ^{2}}\,dW_{1}^{Q^{X}}(t)+\rho
\,dW_{2}^{Q^{X}}(t)\right) ,\,  \label{hes5} \\
d\frac{1}{\sqrt{f}} &=&\left( r_{X}(t)-r_{\$}\right) \frac{1}{\sqrt{f}}dt-%
\frac{\sqrt{v}}{2}\frac{1}{\sqrt{f}}\left( \sqrt{1-\rho ^{2}}%
\,dW_{1}^{Q^{X}}(t)+\rho \,dW_{2}^{Q^{X}}(t)\right) ,  \notag \\
dv &=&\kappa \,(\theta -v)dt+\delta \sqrt{v}dW_{2}^{Q^{X}},\ v(0)=v_{0}, 
\notag
\end{eqnarray}%
where, as before (see (\ref{eq:rxxx})), 
\begin{equation}
r_{X}(t)=\frac{r_{\$}+r_{\text{\euro }}}{2}+\frac{v(t)}{8}  \label{hes_rxxx}
\end{equation}%
and, without changing the notation, the new $\kappa $ and $\theta $ in (\ref%
{hes5}) are equal to $\kappa +\lambda \delta $ and $\kappa \theta /(\kappa
+\lambda \delta )$, respectively, in terms of the old $\kappa $ and $\theta $
from (\ref{hes1}). Then (see Theorem~\ref{thm:priceQX1}), e.g. the price of
the call (in USD) for buying \euro $1$ for \$$K$ is equal to%
\begin{equation}
C_{\text{\euro }/\$}(0,T,K)=\sqrt{f(0)}\mathbb{E}_{Q^{X}}\left[
D_{X}(0,T)\left( \sqrt{f(T)}-\frac{K}{\sqrt{f(T)}}\right) _{+}\right] ,
\label{hes6}
\end{equation}%
where $\sqrt{f(T)}$ and $1/\sqrt{f(T)}$ are from (\ref{hes5}).

Now we will rewrite (\ref{hes5}) under the T-forward measure $Q_{T}^{X}$
using the results of Section~\ref{sec:QXT}. By (\ref{hes_rxxx}), we have 
\begin{eqnarray*}
P_{X}(t,T) &=&\mathbb{E}_{Q^{X}}\left[ D_{X}(t,T)|{\mathcal{F}}_{t}\right] \\
&=&\mathbb{E}_{Q^{X}}\left[ \left. \exp \left(
-\int_{t}^{T}r_{X}(s)ds\right) \right\vert {\mathcal{F}}_{t}\right] \\
&=&\exp \left( -\frac{r_{\$}+r_{\text{\euro }}}{2}(T-t)\right) \mathbb{E}%
_{Q^{X}}\left[ \left. \exp \left( -\int_{t}^{T}\frac{v(s)}{8}ds\right)
\right\vert v(t)\right] .
\end{eqnarray*}

The stochastic X short rate $r_{X}(t)$ defined by (\ref{hes_rxxx}) with $%
v(t) $ from (\ref{hes5}) possesses an affine term structure (see e.g. \cite%
{Bjork}): 
\begin{equation}
P_{X}(t,T)=\exp \left( -\frac{r_{\$}+r_{\text{\euro }}}{2}%
(T-t)+A(T-t)-C(T-t)v(t)\right) ,  \label{hes7}
\end{equation}%
where 
\begin{eqnarray*}
A(t) &=&\frac{2\kappa \theta }{\delta ^{2}}\ln \left( \frac{2\beta e^{(\beta
+\kappa )t/2}}{(\beta +\kappa )\left( e^{\beta t}-1\right) +2\beta }\right) ,
\\
C(t) &=&\frac{1}{4}\frac{e^{\beta t}-1}{(\beta +\kappa )\left( e^{\beta
t}-1\right) +2\beta },
\end{eqnarray*}%
with 
\begin{equation*}
\beta =\sqrt{\kappa ^{2}+\delta ^{2}/4}.
\end{equation*}%
We note that 
\begin{equation*}
dP_{X}=r_{X}(t)P_{X}dt-\delta C(T-t)\sqrt{v}P_{X}dW_{2}^{Q^{X}}(t).
\end{equation*}%
Then we obtain 
\begin{eqnarray}
\mathbb{E}_{Q^{X}}\left[ \left. \frac{dQ_{T}^{X}}{dQ^{X}}\right\vert 
\mathcal{F}_{t}\right] &=&\frac{P_{X}(t,T)}{P_{X}(0,T)B_{_{X}}(t)}
\label{hes71} \\
&=&\exp \left( -\frac{1}{2}\int_{0}^{t}C^{2}(T-s)\delta
^{2}v(s)ds-\int_{0}^{t}C(T-s)\delta \sqrt{v(s)}dW_{2}^{Q^{X}}(s)\right) . 
\notag
\end{eqnarray}%
Hence 
\begin{equation*}
dW_{2}^{Q_{T}^{X}}=dW_{2}^{Q^{X}}+C(T-t)\delta \sqrt{v(t)}dt.
\end{equation*}%
To complete the change of measure, we need to look at $W_{1}^{Q_{T}^{X}}(t)$%
. To this end, we recall that both forward EUR-X and USD-X exchange rates, 
\begin{equation*}
F_{\text{\euro }/X}(t,T)=e^{-r_{\text{\euro }}(T-t)}\frac{\sqrt{f(t)}}{%
P_{X}(t,T)}
\end{equation*}%
and 
\begin{equation*}
F_{\$/X}(t,T)=e^{-r_{\$}(T-t)}\frac{1}{P_{X}(t,T)\sqrt{f(t)}},
\end{equation*}%
should be $Q_{T}^{X}$-martingales. It is not difficult to check that to
achieve the above no-arbitrage requirement, we need%
\begin{equation*}
dW_{1}^{Q^{X}}(t)=dW_{1}^{Q_{T}^{X}}(t),
\end{equation*}%
which is natural since the change of measure (\ref{hes71}) does not depend
on $W_{1}^{Q^{X}}(t).$

Thus, applying Theorem~\ref{thm:main} to the Heston model setting, we can
price, e.g. the call option as (see (\ref{eq:thm1})): 
\begin{eqnarray}
C_{\text{\euro }/\$}(0,T,K) &=&\frac{e^{-r_{\$}T}}{\mathbb{E}_{Q_{T}^{X}}%
\left[ \frac{1}{\sqrt{f(T)}}\right] }\mathbb{E}_{Q_{T}^{X}}\left[ \left( 
\sqrt{f(T)}-\frac{K}{\sqrt{f(T)}}\right) _{+}\right]   \label{hes8} \\
&=&\frac{e^{-r_{\$}T}}{\mathbb{E}_{Q_{T}^{X}}\left[ F_{\$/X}(T,T)\right] }%
\mathbb{E}_{Q_{T}^{X}}\left[ \left( F_{\text{\euro }/X}(T,T)-KF_{\$/X}(T,T)%
\right) _{+}\right] ,  \notag
\end{eqnarray}%
and 
\begin{equation}
C_{\$/\text{\euro }}(0,T,K)=\frac{e^{-r_{\text{\euro }}T}}{\mathbb{E}%
_{Q_{T}^{X}}\left[ F_{\text{\euro }/X}(T,T)\right] }\mathbb{E}_{Q_{T}^{X}}%
\left[ \left( F_{\$/X}(T,T)-KF_{\text{\euro }/X}(T,T)\right) _{+}\right] ,
\label{hes89}
\end{equation}%
where 
\begin{eqnarray}
dF_{\text{\euro }/X}(t,T) &=&\frac{\sqrt{v}}{2}F_{\text{\euro }%
/X}(t,T)\left( \sqrt{1-\rho ^{2}}\,dW_{1}^{Q^{X}}(t)+\rho
\,dW_{2}^{Q_{T}^{X}}\right)   \label{hes9} \\
&&+\delta C(T-t)\sqrt{v}F_{\text{\euro }/X}(t,T)dW_{2}^{Q_{T}^{X}},  \notag
\\
dF_{\$/X}(t,T) &=&-\frac{\sqrt{v}}{2}F_{\$/X}(t,T)\left( \sqrt{1-\rho ^{2}}%
\,dW_{1}^{Q^{X}}(t)+\rho \,dW_{2}^{Q_{T}^{X}}\right)   \notag \\
&&+\delta C(T-t)\sqrt{v}F_{\$/X}(t,T)dW_{2}^{Q_{T}^{X}},  \notag \\
dv &=&\left( \kappa -C(T-t)\delta ^{2}\right) \left( \frac{\kappa \theta }{%
\kappa -C(T-t)\delta ^{2}}-v\right) dt+\delta \sqrt{v}dW_{2}^{Q_{T}^{X}},\
v(0)=v_{0},  \notag
\end{eqnarray}%
and we require that for $0\leq t\leq T$%
\begin{equation}
\kappa /\delta ^{2}>C(t).  \label{hes10}
\end{equation}

We note that in comparison with the classical Heston model (\ref{hes1}), the
model (\ref{hes9}) has time dependence in the coefficients. For other
time-dependent Heston models, see e.g. \cite{Gobet10,Oost11} and references
therein.

\subsection{SABR model\label{sec:sabr}}

For simplicity again, let the interest rates for the USD and EUR money
markets, $r_{\$}$ and $r_{\text{\euro }}$, be constant. Following Section~%
\ref{sec:QX}, we can re-write the classical SABR model \cite%
{hagan2002managing} for EUR-USD exchange rate $f(t)$ under the measure $Q^{X}
$ and the corresponding SDEs for $S_{\text{\euro }/X}=\sqrt{f}$ and $S_{%
\text{\$}/X}=1/\sqrt{f}$ take the form 
\begin{eqnarray}
d\sqrt{f} &=&\left( r_{X}(t)-r_{\text{\euro }}\right) \sqrt{f}dt+\frac{%
\sigma (t)}{2}\sqrt{f}\left( \sqrt{1-\rho ^{2}}\,dW_{1}^{Q^{X}}(t)+\rho
\,dW_{2}^{Q^{X}}(t)\right) ,  \label{sabr0} \\
d\frac{1}{\sqrt{f}} &=&\left( r_{X}(t)-r_{\$}\right) \frac{1}{\sqrt{f}}dt-%
\frac{\sigma (t)}{2}\frac{1}{\sqrt{f}}\left( \sqrt{1-\rho ^{2}}%
\,dW_{1}^{Q^{X}}(t)+\rho \,dW_{2}^{Q^{X}}(t)\right) ,  \notag
\end{eqnarray}%
\begin{equation}
d\sigma =\nu \sigma dW_{2}^{Q^{X}}(t),\ \sigma (0)=\alpha ,  \label{sabr2}
\end{equation}%
where $W_{1}^{Q^{X}}(t)$ and $W_{2}^{Q^{X}}$ are independent standard Wiener
processes under $Q^{X}$, $\rho \in (-1,0]$ is the correlation coefficient, $%
\nu >0$ is the volatility of the volatility $\sigma (t)$, $\alpha $ is a
positive constant, and (see (\ref{eq:rxxx})) 
\begin{equation*}
r_{X}(t)=\frac{r_{\$}+r_{\text{\euro }}}{2}+\frac{\sigma ^{2}(t)}{8}\ .
\end{equation*}%
Note that the parameter known as $\beta $ in the classical SABR model is
taken to be equal to $1$ here, which is the typical requirement for FX
modelling as it ensures that the SDE for the exchange rate for the inverse
pair $1/f$ has the same form as for $f.$

By Theorem~\ref{thm:priceQX1}, e.g. the price of the call (in USD) for
buying \euro $1$ for \$$K$ is equal to 
\begin{equation}
C_{\text{\euro }/\$}((0,T,K)=\sqrt{f(0)}\mathbb{E}_{Q^{X}}\left[
D_{X}(0,T)\left( \sqrt{f(T)}-\frac{K}{\sqrt{f(T)}}\right) _{+}\right] ,
\label{sabr3}
\end{equation}%
where $\sqrt{f(T)}$ and $1/\sqrt{f(T)}$ satisfy (\ref{sabr0}), (\ref{sabr2}).

\subsection{Extended skew normal model\label{sec:esn}}

In this section we consider another illustration of Theorem~\ref{thm:main}.
Here we start not with a model under a `market' measure but with a direct
assumption on the distribution of the exchange rate under a forward measure $%
Q_{T}^{X}$ on the intermediate market.

We assume that under a T-forward measure $Q_{T}^{X}$ the EUR-USD exchange
rate $f(T)$ can be written as 
\begin{equation}
f(T)=\bar{F}e^{Z},  \label{eq:distrf}
\end{equation}%
where $\bar{F}>0$ is a constant and $Z$ is a random variable such that $%
\mathbb{E}\left[ e^{Z}\right] $ exists and the no-arbitrage condition (\ref%
{eq:arbit}) is satisfied by $f(T)$. Here the no-arbitrage condition (\ref%
{eq:arbit}) implies that 
\begin{equation}
\bar{F}=F\,\frac{\mathbb{E}\left[ e^{-Z/2}\right] }{\mathbb{E}\left[ e^{Z/2}%
\right] },  \label{eq:fbar}
\end{equation}%
where we neglect the full notation $\mathbb{E}_{Q_{T}^{X}}[\cdot ]$ and
write $\mathbb{E}[\cdot ]$ instead as in this section we work with the
measure $Q_{T}^{X}$ only and we also write here $F$ instead of $F_{\text{%
\euro }/\$}(0,T)$ for the current forward EUR-USD exchange rate (see (\ref%
{eq:forward})). We use this simplified notation throughout this section,
which should not cause any confusion. The interest rates for the USD and EUR
money markets, $r_{\$}$ and $r_{\text{\euro }}$, are assumed to be constant.

Further, (\ref{eq:thm1}) (i.e. Theorem~\ref{thm:main}), (\ref{eq:distrf})
and (\ref{eq:fbar}) imply that the price (in USD) of the European call to
buy \euro $1$ for $\$K$ at the maturity $T$ is%
\begin{align}
C_{\text{\euro }/\$}(0,T,K)& =\frac{e^{-r_{\$}T}}{\mathbb{E}\left[ \frac{1}{%
\sqrt{f(T)}}\right] }\mathbb{E}\left[ \left( \sqrt{f(T)}-\frac{K}{\sqrt{f(T)}%
}\right) _{+}\right]   \label{eq:call4} \\
& =e^{-r_{\$}T}\,\frac{\sqrt{\bar{F}}}{\mathbb{E}\left[ e^{-Z/2}\right] }%
\mathbb{E}\left[ \left( \sqrt{f(T)}-\frac{K}{\sqrt{f(T)}}\right) \mathbbm{1}%
_{\bar{F}e^{Z}>K}\right]   \notag \\
& =e^{-r_{\$}T}\,\frac{\sqrt{\bar{F}}}{\mathbb{E}\left[ e^{-Z/2}\right] }%
\left( \sqrt{\bar{F}}\,\mathbb{E}\left[ e^{Z/2}\mathbbm{1}_{Z>z_{0}}\right] -%
\frac{K}{\sqrt{\bar{F}}}\,\mathbb{E}\left[ e^{-Z/2}\mathbbm{1}_{Z>z_{0}}%
\right] \right)   \notag \\
& =e^{-r_{\$}T}\,\left( \bar{F}\,\frac{\mathbb{E}\left[ e^{Z/2}\mathbbm{1}%
_{Z>z_{0}}\right] }{\mathbb{E}\left[ e^{-Z/2}\right] }-K\,\frac{\mathbb{E}%
\left[ e^{-Z/2}\mathbbm{1}_{Z>z_{0}}\right] }{\mathbb{E}\left[ e^{-Z/2}%
\right] }\right)   \notag \\
& =e^{-r_{\$}T}\,\left( F\,\frac{\mathbb{E}\left[ e^{Z/2}\mathbbm{1}%
_{Z>z_{0}}\right] }{\mathbb{E}\left[ e^{Z/2}\right] }-K\,\frac{\mathbb{E}%
\left[ e^{-Z/2}\mathbbm{1}_{Z>z_{0}}\right] }{\mathbb{E}\left[ e^{-Z/2}%
\right] }\right)   \notag \\
& =e^{-r_{\$}T}\,\left( F\,\frac{M(1/2,z_{0})}{M(1/2)}-K\,\frac{M(-1/2,z_{0})%
}{M(-1/2)}\right) ,  \notag
\end{align}%
where $z_{0}=\log \left( K/\bar{F}\right) $ and 
\begin{equation}
M(t)=\mathbb{E}[e^{tZ}]\text{ and }M(t,z_{0})=\mathbb{E}\left[ e^{tZ}%
\mathbbm{1}_{Z>z_{0}}\right] ,  \label{eq:mgm1}
\end{equation}%
which are the moment generating function (MGF) and the restricted MGF for $Z$%
, respectively. Analogous to (\ref{eq:call4}), we can derive the pricing
formulas for the put and also for the call and put for the inverse pair: 
\begin{align*}
P_{\text{\euro }/\$}(0,T,K)& =\frac{e^{-r_{\$}T}}{\mathbb{E}\left[ \frac{1}{%
\sqrt{f(t)}}\right] }\mathbb{E}\left[ \left( \frac{K}{\sqrt{f(T)}}-\sqrt{f(T)%
}\right) _{+}\right]  \\
& =e^{-r_{\$}T}\,\left( K\,\frac{M^{\ast }(-1/2,z_{0})}{M(-1/2)}-F\,\frac{%
M^{\ast }(1/2,z_{0})}{M(1/2)}\right) , \\
C_{\$/\text{\euro }}\left( 0,T,\frac{1}{K}\right) & =\frac{e^{-r_{\text{%
\euro }}T}}{\mathbb{E}\left[ \sqrt{f(t)}\right] }\mathbb{E}\left[ \left( 
\frac{1}{\sqrt{f(T)}}-\frac{\sqrt{f(T)}}{K}\right) _{+}\right]  \\
& =e^{-r_{\text{\euro }}T}\,\left( \frac{1}{F}\,\frac{M^{\ast }(-1/2,z_{0})}{%
M(-1/2)}-\frac{1}{K}\,\frac{M^{\ast }(1/2,z_{0})}{M(1/2)}\right) , \\
P_{\$/\text{\euro }}\left( 0,T,\frac{1}{K}\right) & =\frac{e^{-r_{\text{%
\euro }}T}}{\mathbb{E}\left[ \sqrt{f(t)}\right] }\mathbb{E}\left[ \left( 
\frac{\sqrt{f(T)}}{K}-\frac{1}{\sqrt{f(T)}}\right) _{+}\right]  \\
& =e^{-r_{\text{\euro }}T}\,\left( \frac{1}{K}\,\frac{M(1/2,z_{0})}{M(1/2)}-%
\frac{1}{F}\,\frac{M(-1/2,z_{0})}{M(-1/2)}\right) ,
\end{align*}%
where 
\begin{equation*}
M^{\ast }(t,z_{0})=\mathbb{E}\left[ e^{tZ}\mathbbm{1}_{Z<z_{0}}\right] .
\end{equation*}

Now we will propose a skew normal model for the random variable $Z$. To this
end, we start by introducing a new random variable $V$, which is a
combination of one normal and two shifted half-normal distributed random
variables: 
\begin{equation}
V:=X+\alpha _{1}\max (\beta _{1}-Y,0)+\alpha _{2}\max (Y-\beta _{2},0),
\label{eq:defV}
\end{equation}%
where $X$ and $Y$ are independent random variables with the standard normal
distribution and $\alpha _{1}$, $\alpha _{2}$, $\beta _{1}$, $\beta _{2}\in 
\mathbb{R}$ are parameters. The parameters $\beta _{1}$ and $\beta _{2}$
describe the support domains of the two half-normal distributions which,
from the modeling prospective, should not overlap. Consequently, we are only
interested in the case: 
\begin{equation*}
0<\beta _{1}\leq \beta _{2}.
\end{equation*}%
Since we use the random variable $Z$ in (\ref{eq:distrf}) similarly to as a
Gaussian random variable is used in the geometric Brownian motion model for $%
f(T)$, we then define it as follows 
\begin{equation}
Z=aV,  \label{eq:defZ}
\end{equation}%
where $a=\sigma \sqrt{T}$ with $\sigma $ having the meaning of volatility
and $T$ of the maturity time. The benefit of using $Z$ instead of a Gaussian
random variable is that $Z$ can have heavier tails and can be successfully
used for describing the volatility smile effect. At the same time, $Z$ still
has a very simple distribution which makes the model (\ref{eq:distrf}), (\ref%
{eq:defV}), (\ref{eq:defZ}) very practical as it allows fast calibration.
Indeed, the MGFs (\ref{eq:mgm1}), which we need for pricing calls and puts
(see (\ref{eq:call4})), can be found analytically for this $Z$. The
corresponding expressions are given in the next proposition.

\begin{proposition}
\label{prop:optionpricingNew}For $0<\beta _{1}\leq \beta _{2}$, the MGF $%
M(t) $ and the restricted MGF $M(t,z_{0})$ from (\ref{eq:mgm1}) are equal to 
\begin{align}
M(t)& =e^{\frac{(at)^{2}}{2}}(N(\beta _{2})-N(\beta _{1}))+e^{\frac{t}{2}%
(ta^{2}(1+\alpha _{2}^{2})-2a\alpha _{2}\beta _{2})}N(ta\alpha _{2}-\beta
_{2})  \label{eq:mgf} \\
& +e^{\frac{t}{2}(ta^{2}(1+\alpha _{1}^{2})+2a\alpha _{1}\beta
_{1})}N(ta\alpha _{1}+\beta _{1}),  \notag \\
M(t,z_{0})& =e^{\frac{(at)^{2}}{2}}N\left( at-\frac{z_{0}}{a}\right) \big(%
N(\beta _{2})-N(\beta _{1})\big)+e^{\frac{t}{2}(ta^{2}(1+\alpha
_{1}^{2})+2a\alpha _{1}\beta _{1})}  \label{eq:mgfres} \\
\times & \left\{ N(ta\alpha _{1}+\beta _{1})-N_{2}\left( ta\alpha _{1}+\beta
_{1},\frac{\frac{z_{0}}{a}-at-\alpha _{1}(\beta _{1}+ta\alpha _{1})}{\sqrt{%
1+\alpha _{1}^{2}}};\frac{-\alpha _{1}}{\sqrt{1+\alpha _{1}^{2}}}\right)
\right\}  \notag \\
& +e^{\frac{t}{2}(ta^{2}(1+\alpha _{2}^{2})-2a\alpha _{2}\beta _{2})}  \notag
\\
& \times \left\{ N(ta\alpha _{2}-\beta _{2})-N_{2}\left( ta\alpha _{2}-\beta
_{2},\frac{\frac{z_{0}}{a}-t+\alpha _{2}(\beta _{2}-ta\alpha _{2})}{\sqrt{%
1+\alpha _{2}^{2}}};\frac{-\alpha _{2}}{\sqrt{1+\alpha _{2}^{2}}}\right)
\right\} ,  \notag
\end{align}%
where $N(\cdot )$ is the cdf of the standard normal distribution and $%
N_{2}(\cdot ,\cdot ;\rho )$ is the cdf of the bivariate normal distribution
with zero mean, unit variance, and correlation $\rho $.
\end{proposition}

\begin{figure}[tbp]
\centering
\includegraphics[scale = 0.15]{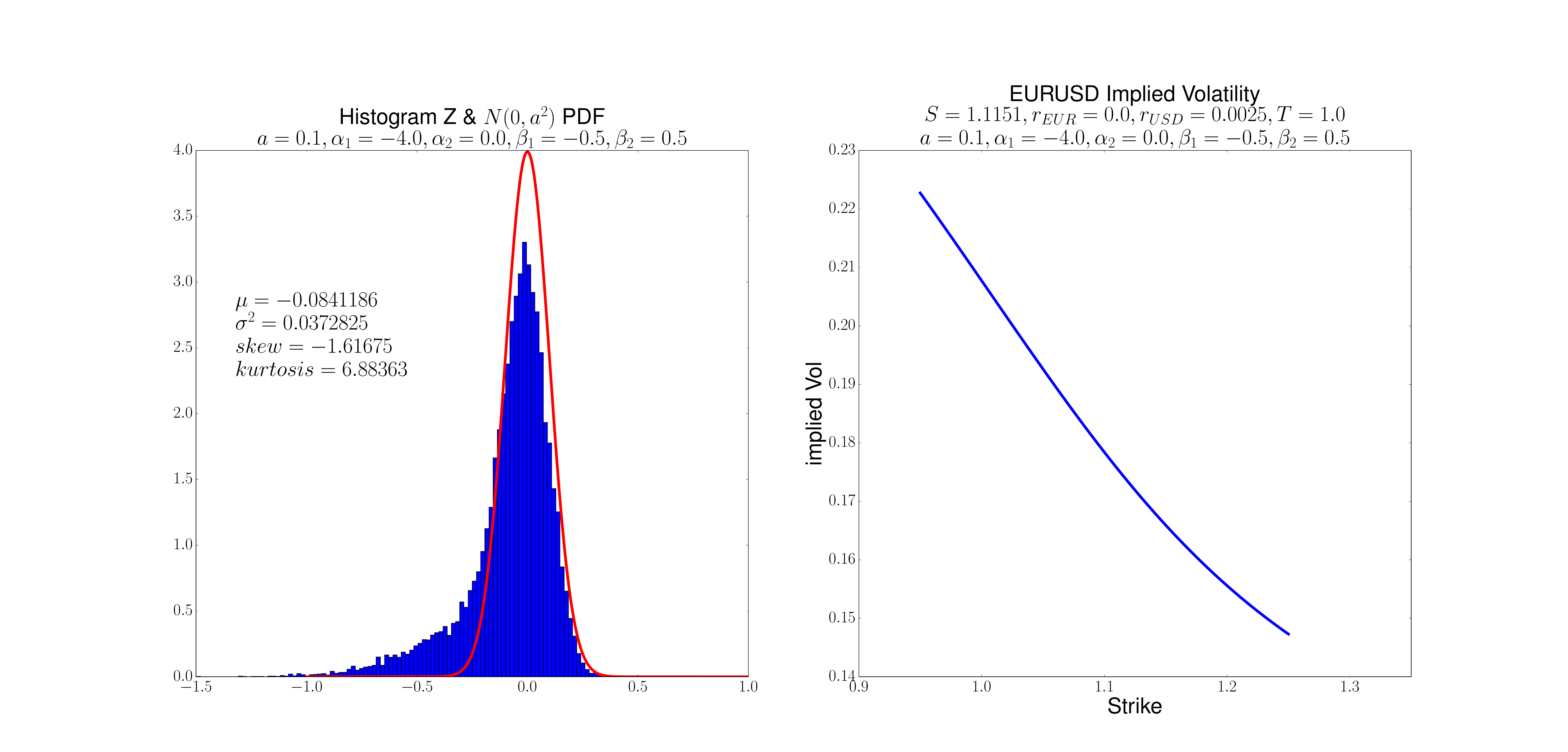}
\caption{Effect of the parameters on the distribution of Z and the
corresponding smile: the case of $\protect\alpha _{1}<0$ and $\protect\alpha %
_{2}=0$.}
\label{fig:HistSmile2}
\end{figure}

\begin{figure}[tbp]
\centering
\includegraphics[scale = 0.15]{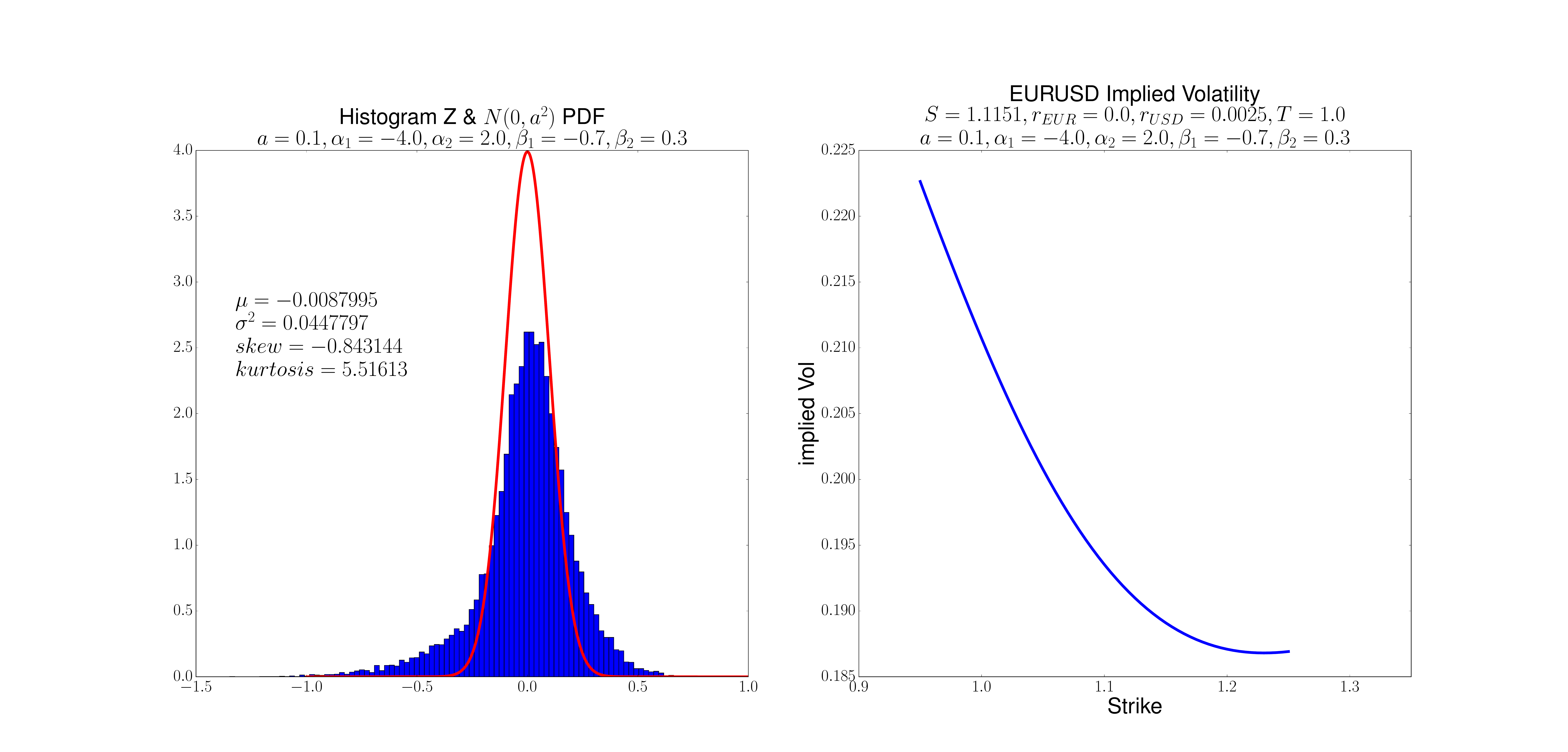}
\caption{ Effect of the parameters on the distribution of Z and the
corresponding smile: the case of $\protect\alpha _{1}<0$ and $\protect\alpha %
_{2}>0$.}
\label{fig:HistSmile4}
\end{figure}

Using the parameters $\alpha _{1}$, $\alpha _{2}$, $\beta _{1}$, and $\beta
_{2}$, we can manipulate with the distribution of $Z$ defined in (\ref%
{eq:defV}), (\ref{eq:defZ}) and, in particular, its skew and kurtosis, which
are equal to 
\begin{eqnarray*}
skew_{Z} &=&\frac{M_{V}^{(3)}(0)-3M_{V}^{(1)}(0)M_{V}^{(2)}(0)+2\left[
M_{V}^{(1)}(0)\right] ^{3}}{\left( M_{V}^{(2)}(0)-\left[ M_{V}^{(1)}(0)%
\right] ^{2}\right) ^{3/2}}, \\
kurtosis_{Z} &=&\frac{M_{V}^{(4)}(0)-4M_{V}^{(1)}(0)M_{V}^{(3)}(0)+6\left[
M_{V}^{(1)}(0)\right] ^{2}M_{V}^{(2)}(0)-3\left[ M_{V}^{(1)}(0)\right] ^{4}}{%
\left( M_{V}^{(2)}(0)-\left[ M_{V}^{(1)}(0)\right] ^{2}\right) ^{2}},
\end{eqnarray*}%
where $M_{V}^{(i)}(0)$ are $i$-th derivatives of the MGF for the random
variable $V$. By putting $\alpha _{1}=\alpha _{2}=0$ in (\ref{eq:defV}), the
random variable $Z$ becomes normal with zero mean and variance $a^{2},$ and
the considered model (\ref{eq:distrf}), (\ref{eq:defV}), (\ref{eq:defZ}) is
reduced to the geometric Brownian motion whose one of the critical
deficiencies is a flat (constant) volatility. In this case $Z$ has $skew=0$
and $kurtosis=3$. In Figure~\ref{fig:HistSmile2}, one can see the difference
of distribution of $Z$ (blue area) compared to a standardized normal
distribution (red line). It can be seen that a parameter set with $\alpha
_{1}<0$ and $\alpha _{2}=0$ results in a bigger left tail in distribution
and a skew in the resulting volatility smile ($skew\approx -1.6$). 
Similarly, in Figure~\ref{fig:HistSmile4}, it can be observed that using $%
\alpha _{1}<0$ to adjust the left tail and $\alpha _{2}>0$ to adjust the
right tail of the smile, we can get an asymmetric distribution and an
asymmetric smile. A smaller $\alpha _{2}$ results in a smaller right tail
and hence in a flatter smile. 
As seen in these figures, by adjusting the parameters $\alpha _{1}$, $\alpha
_{2}$, $\beta _{1}$, $\beta _{2}$, the shape of the distribution and of the
smile can be changed in various ways and it can be associated with the
resulting skew and kurtosis of the log exchange rate. Hence, after
calibrating the parameters of $Z$ to FX market data, we can compare $%
skew_{Z} $ with zero skew and $kurtosis_{Z}$ with the kurtosis of $3$ in the
geometric Brownian motion case and make a conclusion about how far
volatility is from a constant.

\subsection{Model-free approach\label{sec:mf}}

The model-free approach to pricing derivatives has become popular in recent
years \cite{Aus11,Fuk12,Aus14,MM18} (see also references therein). The main
idea of the approach is to construct a density or distribution function of
risky assets under a risk-neutral measure using observed prices of
plain-vanilla options. For clarity of the exposition how this approach works
within our intermediate currency framework, we start with the case of two
currencies. Then we will extend the consideration to the three-currencies
case where we will exploit ideas from \cite{Aus11} (see also \cite{Aus14}).

In this section we will work under a T--forward measure $Q_{T}^{X}.$ Assume
that we know prices of call options $C_{\text{\euro }/\$}(0;K)$ for all
strikes $K>0$ and let $\rho (x;T)$ be the density of the EUR-USD exchange
rate $f(T)$ under $Q_{T}^{X}.$ According to (\ref{eq:QT2}), we have 
\begin{eqnarray}
V_{\text{\euro }/\$}(0) &=&\frac{e^{-r_{\$}T}}{\mathbb{E}_{Q_{T}^{X}}\left[ 
\sqrt{f(T)}\right] }\mathbb{E}_{Q_{T}^{X}}\left[ g\left( \sqrt{f(T)};\frac{K%
}{\sqrt{f(T)}}\right) \right]  \label{mf1} \\
&=&\frac{e^{-r_{\$}T}}{\mathbb{E}_{Q_{T}^{X}}\left[ \sqrt{f(T)}\right] }%
\int_{0}^{\infty }\frac{1}{\sqrt{x}}g\left( x;K\right) \rho (x;T)dx  \notag
\\
&=&\int_{0}^{\infty }g\left( x;K\right) \frac{\partial ^{2}}{\partial x^{2}}%
C_{\text{\euro }/\$}(0;x)dx,  \notag
\end{eqnarray}%
as simple calculations give%
\begin{equation}
\frac{e^{-r_{\$}T}}{\sqrt{K}\mathbb{E}_{Q_{T}^{X}}\left[ \sqrt{f(T)}\right] }%
\rho (K;T)=\frac{\partial ^{2}}{\partial K^{2}}C_{\text{\euro }/\$}(0;K).
\label{mf2}
\end{equation}%
Typically, observed data are expressed via volatility smile data $\sigma (K)$
and from (\ref{e1}) we have 
\begin{equation}
C_{\text{\euro }/\$}(0;K)=F_{\text{\euro }/\$}e^{-r_{\$}T}N\left( \frac{\log 
\frac{F_{\text{\euro }/\$}}{K}+\sigma ^{2}(K)T/2}{\sigma (K)\sqrt{T}}\right)
-Ke^{-r_{\$}T}N\left( \frac{\log \frac{F_{\text{\euro }/\$}}{K}-\sigma
^{2}(K)T/2}{\sigma (K)\sqrt{T}}\right) .  \label{mf3}
\end{equation}%
Combining (\ref{mf1}) with (\ref{mf3}) and given $\sigma (K)$, we can price
any FX derivative $V_{\text{\euro }/\$}(0)$ and analogously any derivative $%
V_{\$/\text{\euro }}(0)$ based on a smile from one of the markets. Note that
the smile data computed from $C_{\text{\euro }/\$}(0;K)$ coincide with smile
data computed from $C_{\$/\text{\euro }}(0;K)$ and that the prices $V_{\text{%
\euro }/\$}(0)$ and $V_{\$/\text{\euro }}(0)$ are consistent with each other
thanks to using the intermediate currency framework. \medskip

Now we progress to the three-currencies case. Let us assume that we are
interested in the GBP-USD-EUR currency triangle, where we denote GBP as
currency 1, USD as currency 2, and EUR as currency 3. As before, the
interest rates for the GBP, USD and EUR money markets, $r_{\pounds },$ $%
r_{\$}$ and $r_{\text{\euro }}$, are assumed to be constant.

Consider a best-of option on the EUR market which payoff is equal to 
\begin{equation}
b(T)=\max \left\{ \frac{\left( S_{\pounds /\text{\euro }}(T)-K_{1}\right)
_{+}}{K_{1}},\frac{\left( S_{\$/\text{\euro }}(T)-K_{2}\right) _{+}}{K_{2}}%
\right\} .  \label{mf4}
\end{equation}%
As it is known \cite{Aus11} and \cite[Sec. 11.7.1]{Aus14}, the value of a
best-of option is arbitrary close to values of plain-vanilla calls on $S_{%
\pounds /\text{\euro }}(T)$ or $S_{\$/\text{\euro }}(T)$ or to a vanilla
option on the cross $S_{\pounds /\$}(T).$ Hence, a model used for FX pricing
should price a best-of option and plain-vanilla options in a consistent
manner.

By (\ref{eq:intermediateExchangeMulti}) we have%
\begin{equation}
S_{\pounds /X}=S_{\pounds /\text{\euro }}^{2/3}(T)S_{\$/\text{\euro }%
}^{-1/3}(T),\ \ S_{\$/X}=S_{\pounds /\text{\euro }}^{-1/3}(T)S_{\$/\text{%
\euro }}^{2/3}(T),\ S_{\text{\euro }/X}=S_{\pounds /\text{\euro }%
}^{-1/3}(T)S_{\$/\text{\euro }}^{-1/3}(T).  \label{mf50}
\end{equation}%
Using the pricing formula (\ref{eq:thm1Multi2}), we get 
\begin{equation}
V_{\text{\euro }}(0)=\frac{e^{-r_{\text{\euro }}T}}{\mathbb{E}_{Q_{T}^{X}}%
\left[ S_{\text{\euro }/X}(T)\right] }\mathbb{E}_{Q_{T}^{X}}G(T)=\frac{%
e^{-r_{\text{\euro }}T}}{\mathbb{E}_{Q_{T}^{X}}\left[ S_{\text{\euro }/X}(T)%
\right] }\mathbb{E}_{Q_{T}^{X}}\left[ S_{\text{\euro }/X}(T)g(T)\right] ,
\label{mf5}
\end{equation}%
where $g(T)$ is an arbitrary payoff on the EUR market. Hence, for the
best-of option we have 
\begin{eqnarray*}
v_{\text{\euro }}(0) &=&\frac{e^{-r_{\text{\euro }}T}}{\mathbb{E}_{Q_{T}^{X}}%
\left[ S_{\text{\euro }/X}(T)\right] }\mathbb{E}_{Q_{T}^{X}}\ \left[ S_{%
\text{\euro }/X}(T)\ \max \left\{ \frac{\left( S_{\pounds /\text{\euro }%
}(T)-K_{1}\right) _{+}}{K_{1}},\frac{\left( S_{\$/\text{\euro }%
}(T)-K_{2}\right) _{+}}{K_{2}}\right\} \right] \\
&=&\frac{e^{-r_{\text{\euro }}T}}{\mathbb{E}_{Q_{T}^{X}}\left[ S_{\text{%
\euro }/X}(T)\right] } \\
&&\times \mathbb{E}_{Q_{T}^{X}}\ \left[ S_{\pounds /\text{\euro }%
}^{-1/3}(T)S_{\$/\text{\euro }}^{-1/3}(T)\ \max \left\{ \frac{\left( S_{%
\pounds /\text{\euro }}(T)-K_{1}\right) _{+}}{K_{1}},\frac{\left( S_{\$/%
\text{\euro }}(T)-K_{2}\right) _{+}}{K_{2}}\right\} \right] \\
&=&\frac{e^{-r_{\text{\euro }}T}}{\mathbb{E}_{Q_{T}^{X}}\left[ S_{\text{%
\euro }/X}(T)\right] } \\
&&\times \int_{0}^{\infty }\int_{0}^{\infty }x^{-1/3}y^{-1/3}\ \max \left\{ 
\frac{\left( x-K_{1}\right) _{+}}{K_{1}},\frac{\left( y-K_{2}\right) _{+}}{%
K_{2}}\right\} \rho (x,y;T)dxdy,
\end{eqnarray*}%
where $\rho (x,y;T)$ is the joint density of the exchange rates $S_{\pounds /%
\text{\euro }}(T)$ and $S_{\$/\text{\euro }}(T)$ under $Q_{T}^{X}.$

Differentiation of $b(T)$ (see \cite{Aus11,Aus14}) gives: 
\begin{equation*}
\left[ 1+K_{1}\frac{\partial }{\partial K_{1}}+K_{2}\frac{\partial }{%
\partial K_{2}}\right] b(T)=\mathbf{I}(S_{\pounds /\text{\euro }%
}(T)<K_{1},S_{\pounds /\text{\euro }}(T)<K_{2})-1.
\end{equation*}%
Therefore 
\begin{equation}
\frac{\partial ^{2}}{\partial K_{1}\partial K_{2}}\left[ 1+K_{1}\frac{%
\partial }{\partial K_{1}}+K_{2}\frac{\partial }{\partial K_{2}}\right] v_{%
\text{\euro }}(0)=\frac{e^{-r_{\text{\euro }}T}}{\mathbb{E}_{Q_{T}^{X}}\left[
S_{\text{\euro }/X}(T)\right] }K_{1}^{-1/3}K_{2}{}^{-1/3}\ \rho
(K_{1},K_{2};T).  \label{mf6}
\end{equation}

Note that we have from (\ref{eq:arbitmulti}) 
\begin{equation}
\frac{\mathbb{E}_{Q_{T}^{X}}\left[ S_{c_{i}/X}(T)\right] }{\mathbb{E}%
_{Q_{T}^{X}}\left[ S_{\text{\euro }/X}(T)\right] }=F_{c_{i}/\text{\euro }%
}(0),\quad i=1,2,  \label{mf67}
\end{equation}%
and from (\ref{eq:thm1Multi2})%
\begin{equation}
V_{c_{N}}(0)=\frac{e^{-r_{N}T}}{\mathbb{E}_{Q_{T}^{X}}\left[ S_{c_{N}/X}(T)%
\right] }\mathbb{E}_{Q_{T}^{X}}\left[ S_{c_{N}/X}(T)g(T)\right] ,
\label{mf68}
\end{equation}%
where $N$ can be any of the three currencies with $g(T)$ being in the
currency $N.$ Therefore (taking also into account (\ref{mf50})), if we can
evaluate (\ref{mf6}) from market data, then we can price any FX derivatives
on any of the three markets using the same $\rho (K_{1},K_{2};T)$ and, thus,
ensuring consistency of FX option pricing across different markets. Note
that in comparison with (\ref{eq:thm1Multi2}) we do not assume in (\ref{mf68}%
) that $g(T)$ is first-order homogeneous.

Market data in the case of three currencies are typically presented via
three volatility smiles: $\sigma _{1}(K)$\ and $\sigma _{2}(K)$ from vanilla
options on GBP-EUR and USD-EUR, respectively, and $\sigma _{3}(K)$ from the
cross, GBP-USD. To compute values on the smile curves from observed option
prices in the context of our intermediate currency approach, the
Garman-Kohlhagen formulas given below for completeness of the exposition
should be used. To complete, the model-free pricing, we need to express the
current price $v(0)$ of the best-of option via the three volatility smiles.
To this end, we need to find $v(0)$ assuming that the exchange rates follow
geometrical Brownian motions under a T--forward measure $Q_{T}^{X},$ which
coincides with the EMM $Q^{X}.$

In accordance with the no-arbitrage condition (\ref{mf67}), we set 
\begin{align}
S_{\pounds /\text{\euro }}(T)& =F_{\pounds /\text{\euro }}\exp \left( -\frac{%
T}{6}\left[ \sigma _{1}^{2}-2\sigma _{1}\sigma _{2}\rho _{12}\right] +\sigma
_{1}\sqrt{T}X_{1}\right) ,  \label{mf7} \\
S_{\$/\text{\euro }}(T)& =F_{\$/\text{\euro }}\exp \left( -\frac{T}{6}\left[
\sigma _{2}^{2}-2\sigma _{1}\sigma _{2}\rho _{12}\right] +\sigma _{2}\sqrt{T}%
X_{2}\right) ,  \notag
\end{align}%
where $F_{\pounds /\text{\euro }}=F_{\pounds /\text{\euro }}(0,T)$ and $%
F_{\$/\text{\euro }}=F_{\$/\text{\euro }}(0,T)$ are the current forward
GBR-EUR and USD-EUR exchange rate, respectively, and $X_{i}\sim N(0,1)$ with
correlation coefficient $\rho _{12}$. Then the GBP-USD exchange rate is
equal to 
\begin{equation}
S_{\pounds /\$}(T)=\frac{S_{\pounds /\text{\euro }}(T)}{S_{\$/\text{\euro }%
}(T)}=F_{\pounds /\$}\exp \left( -\frac{T}{6}\left[ \sigma _{3}^{2}-2\sigma
_{2}\sigma _{3}\rho _{23}\right] +\sigma _{3}\sqrt{T}X_{3}\right) ,
\label{mf8}
\end{equation}%
where 
\begin{equation*}
F_{\pounds /\$}=\frac{F_{\pounds /\text{\euro }}}{F_{\$/\text{\euro }}},\
\sigma _{3}^{2}=\sigma _{1}^{2}-2\sigma _{1}\sigma _{2}\rho _{12}+\sigma
_{2}^{2}.
\end{equation*}%
and $X_{3}\sim N(0,1)$ with the correlation coefficients 
\begin{equation*}
\rho _{13}=\frac{\sigma _{1}^{2}+\sigma _{3}^{2}-\sigma _{2}^{2}}{2\sigma
_{1}\sigma _{3}},\ \ \rho _{23}=\frac{\sigma _{2}^{2}+\sigma _{3}^{2}-\sigma
_{1}^{2}}{2\sigma _{2}\sigma _{3}}
\end{equation*}%
with $X_{1}$ and $X_{2},$ respectively.

We have 
\begin{eqnarray*}
\frac{S_{\text{\euro }/X}(T)}{\mathbb{E}_{Q_{T}^{X}}\left[ S_{\text{\euro }%
/X}(T)\right] } &=&\frac{\exp \left( -\frac{1}{3}\sigma _{1}\sqrt{T}X_{1}-%
\frac{1}{3}\sigma _{2}\sqrt{T}X_{2}\right) }{\mathbb{E}_{Q_{T}^{X}}\left[
\exp \left( -\frac{1}{3}\sigma _{1}\sqrt{T}X_{1}-\frac{1}{3}\sigma _{2}\sqrt{%
T}X_{2}\right) \right] } \\
&=&\exp \left( -\frac{T}{18}\left[ \sigma _{1}^{2}+2\sigma _{1}\sigma
_{2}\rho +\sigma _{2}^{2}\right] -\frac{1}{3}\sigma _{1}\sqrt{T}X_{1}-\frac{1%
}{3}\sigma _{2}\sqrt{T}X_{2}\right) ,
\end{eqnarray*}%
and it is not difficult to show that the above expression is the
Radon-Nikodym derivative $\dfrac{dQ_{T}^{\text{\euro }}}{dQ_{T}^{X}}$ of the
T--forward measure $Q_{T}^{\text{\euro }}$ on the EUR market with respect to 
$Q_{T}^{X}.$ Then 
\begin{equation*}
V_{\text{\euro }}(0)=\frac{e^{-r_{\text{\euro }}T}}{\mathbb{E}_{Q_{T}^{X}}%
\left[ S_{\text{\euro }/X}(T)\right] }\mathbb{E}_{Q_{T}^{X}}\left[ S_{\text{%
\euro }/X}(T)g(T)\right] =e^{-r_{\text{\euro }}T}\mathbb{E}_{Q_{T}^{\text{%
\euro }}}\left[ g(T)\right] .
\end{equation*}%
Hence, the corresponding Garman-Kohlhagen formulas for calls are given by
(see, e.g. \cite{Bjork}): 
\begin{equation*}
C_{\pounds /\text{\euro }}(0;K)=F_{\pounds /\text{\euro }}e^{-r_{\text{\euro 
}}T}N\left( \frac{\ln (F_{\pounds /\text{\euro }}/K)+\sigma _{1}^{2}T/2}{%
\sigma _{1}\sqrt{T}}\right) -Ke^{-r_{\text{\euro }}T}N\left( \frac{\ln (F_{%
\pounds /\text{\euro }}/K)-\sigma _{1}^{2}T/2}{\sigma _{1}\sqrt{T}}\right) ,
\end{equation*}%
\begin{equation*}
C_{\$/\text{\euro }}(0;K)=F_{\$/\text{\euro }}e^{-r_{\text{\euro }}T}N\left( 
\frac{\ln (F_{\$/\text{\euro }}/K)+\sigma _{2}^{2}T/2}{\sigma _{2}\sqrt{T}}%
\right) -Ke^{-r_{\text{\euro }}T}N\left( \frac{\ln (F_{\$/\text{\euro }%
}/K)-\sigma _{2}^{2}T/2}{\sigma _{2}\sqrt{T}}\right) .
\end{equation*}%
Analogously 
\begin{eqnarray*}
C_{\pounds /\$}(0;K) &=&\frac{e^{-r_{\text{\pounds }}T}}{\mathbb{E}%
_{Q_{T}^{X}}\left[ S_{\text{\pounds }/X}(T)\right] }\mathbb{E}_{Q_{T}^{X}}%
\left[ S_{\text{\pounds }/X}(T)(S_{\$/\text{\pounds }}(T)-K)_{+}\right] \\
&=&e^{-r_{\text{\pounds }}T}\mathbb{E}_{Q_{T}^{\text{\pounds }}}\left[
(S_{\$/\text{\pounds }}(T)-K)_{+}\right] \\
&=&F_{\pounds /\$}e^{-r_{\$}T}N\left( \frac{\ln (F_{\pounds /\$}/K)+\sigma
_{3}^{2}T/2}{\sigma _{3}\sqrt{T}}\right) -Ke^{-r_{\$}T}N\left( \frac{\ln (F_{%
\pounds /\$}/K)-\sigma _{3}^{2}T/2}{\sigma _{3}\sqrt{T}}\right) .
\end{eqnarray*}%
We also have (see \cite{Mar78,Stu82,Aus11}): 
\begin{eqnarray}
v_{\text{\euro }}(0) &=&\frac{e^{-r_{\text{\euro }}T}}{\mathbb{E}_{Q_{T}^{X}}%
\left[ S_{\text{\euro }/X}(T)\right] }  \label{mf11} \\
&&\times \mathbb{E}_{Q_{T}^{X}}\ \left[ S_{\text{\euro }/X}(T)\ \max \left\{ 
\frac{\left( S_{\pounds /\text{\euro }}(T)-K_{1}\right) _{+}}{K_{1}},\frac{%
\left( S_{\$/\text{\euro }}(T)-K_{2}\right) _{+}}{K_{2}}\right\} \right] 
\notag \\
&=&e^{-r_{\text{\euro }}T}\mathbb{E}_{Q_{T}^{\text{\euro }}}\ \left[ \max
\left\{ \frac{\left( S_{\pounds /\text{\euro }}(T)-K_{1}\right) _{+}}{K_{1}},%
\frac{\left( S_{\$/\text{\euro }}(T)-K_{2}\right) _{+}}{K_{2}}\right\} %
\right]  \notag \\
&=&e^{-r_{\text{\euro }}T}\left[ \frac{F_{\pounds /\text{\euro }}}{K_{1}}%
N(d_{1}^{+},d_{3}^{+};\rho _{13})+\frac{F_{\$/\text{\euro }}}{K_{2}}%
N(d_{2}^{+},d_{3}^{-};\rho _{23})+N(-d_{1}^{-},-d_{2}^{-};\rho _{12})-1%
\right] ,  \notag
\end{eqnarray}%
where 
\begin{equation*}
d_{i}^{\pm }=\frac{\ln (F_{i}/K_{i})\pm \sigma _{i}^{2}T/2}{\sigma _{i}\sqrt{%
T}}
\end{equation*}%
and $F_{1}=F_{\pounds /\text{\euro }},$ $F_{2}=F_{\$/\text{\euro }},$ and $%
F_{3}=F_{\pounds /\$}.$

Now we put implied volatility smiles $\sigma _{i}(K_{i}),$ $i=1,2,3,$ with $%
K_{3}=K_{1}/K_{2}$ in (\ref{mf11}) and evaluate the left-hand side of (\ref%
{mf6}). As a result, we obtain for $v_{\text{\euro }}(0)=v_{\text{\euro }%
}(0;K_{1},K_{2},\sigma _{1}(K_{1}),\sigma _{2}(K_{2}),\allowbreak \sigma
_{3}(K_{1}/K_{2}))$ from (\ref{mf11}) (see \cite[Ch. 11]{Aus14}):%
\begin{eqnarray}
U(K_{1},K_{2}) &:&=\left[ 1+K_{1}\frac{\partial }{\partial K_{1}}+K_{2}\frac{%
\partial }{\partial K_{2}}\right] v_{\text{\euro }}  \label{mf12} \\
&=&e^{-r_{\text{\euro }}T}\left( N(-d_{1}^{-},-d_{2}^{-};\rho _{12})+\left[
K_{1}\sigma _{1}^{\prime }(K_{1})\frac{\partial }{\partial \sigma _{1}}%
+K_{2}\sigma _{2}^{\prime }(K_{2})\frac{\partial }{\partial \sigma _{2}}%
\right] v_{\text{\euro }}\right)  \notag \\
&=&e^{-r_{\text{\euro }}T}\left[ N(-d_{1}^{-},-d_{2}^{-};\rho _{12})+K_{1}%
\sqrt{T}\sigma _{1}^{\prime }(K_{1})N^{\prime }(d_{1}^{-})N\left( \frac{%
d_{1}^{-}\rho _{12}-d_{2}^{-}}{\sqrt{1-\rho _{12}^{2}}}\right) \right. 
\notag \\
&&\left. +K_{2}\sqrt{T}\sigma _{2}^{\prime }(K_{2})N^{\prime
}(d_{2}^{-})N\left( \frac{d_{2}^{-}\rho _{12}-d_{1}^{-}}{\sqrt{1-\rho
_{12}^{2}}}\right) -1\right] .  \notag
\end{eqnarray}

Let us recall how the model-free approach is used in practice: (i) for
observed plain-vanilla prices, compute values of the implied volatilities $%
\sigma _{i}(K_{i})$ by inverting the Garman-Kohlhagen formulas; (ii)
smoothly interpolate the implied values to obtain three smiles $\sigma
_{i}(K_{i});$ (iii) plug-in the smiles in (\ref{mf12}); (iv) use $%
U(K_{1},K_{2})$ (cf. (\ref{mf6}) and (\ref{mf12})) together with (\ref{mf67}%
) to price options on all the three markets by the pricing formula (\ref%
{mf68}). The step (iv) can be either realized via integration by parts (see
Example~\ref{ex:basketpricing2}\ below) or by further differentiation to get 
\begin{equation*}
\dfrac{e^{-r_{\text{\euro }}T}}{\mathbb{E}_{Q_{T}^{X}}\left[ S_{\text{\euro }%
/X}(T)\right] }K_{1}^{-1/3}K_{2}{}^{-1/3}\ \rho (K_{1},K_{2};T)=\frac{%
\partial ^{2}}{\partial K_{1}\partial K_{2}}U(K_{1},K_{2}).
\end{equation*}

We emphasise that thanks to the intermediate currency approach we can
consistently price products for all the six pairs based on a single
calibration.

We remark that the no arbitrage condition imposes the following asymptotic
requirements on smiles \cite{Lee04,Aus11,Aus14}: 
\begin{equation}  \label{eq:mf-req1}
\sigma _{i}^{2}(K)=o(|\ln K|)\text{\ as }K\rightarrow 0,~\infty .
\end{equation}%
Also, to ensure that $-1<\rho _{ij}(K_{1},K_{1})<1$ the smiles should
satisfy \cite{Aus11,Aus14}: 
\begin{eqnarray}  \label{eq:mf-req2}
\sigma _{1}(K_{1})+\sigma _{2}(K_{2}) &>&\sigma _{3}(K_{1}/K_{2}),  \notag \\
\sigma _{2}(K_{2})+\sigma _{3}(K_{1}/K_{2}) &>&\sigma _{1}(K_{1}), \\
\sigma _{1}(K_{1})+\sigma _{3}(K_{1}/K_{2}) &>&\sigma _{2}(K_{2}).  \notag
\end{eqnarray}

\begin{example}
\label{ex:basketpricing2} Consider basket pricing as in Example~\ref%
{ex:basketpricing}. Doing integration by parts twice, we get \cite[Ch. 11]%
{Aus14}: 
\begin{eqnarray}
BasketOption_{\text{\euro }}(0) &=&\frac{e^{-r_{\text{\euro }}T}}{\mathbb{E}%
_{Q_{T}^{X}}\left[ S_{_{\text{\euro }}/X}(T)\right] }  \label{mf14} \\
&&\times \mathbb{E}_{Q_{T}^{X}}\left[ S_{\pounds /\text{\euro }%
}^{-1/3}(T)S_{\$/\text{\euro }}^{-1/3}(T)\left( K-\omega _{1}S_{\pounds /%
\text{\euro }}(T)-\omega _{2}S_{\$/\text{\euro }}(T)\right) _{+}\right] 
\notag \\
&=&\int_{0}^{\infty }\int_{0}^{\infty }\left( K-\omega _{1}x-\omega
_{2}y\right) _{+}\frac{\partial ^{2}}{\partial x\partial y}U(x,y)dxdy  \notag
\\
&=&\int_{0}^{K}U\left( \frac{z}{\omega _{1}},\frac{K-z}{\omega _{2}}\right)
dz.  \notag
\end{eqnarray}%
We can also obtain 
\begin{eqnarray}
BasketOption_{\pounds }(0) &=&\frac{e^{-r_{\pounds }T}}{\mathbb{E}%
_{Q_{T}^{X}}\left[ S_{\pounds /X}(T)\right] }\mathbb{E}_{Q_{T}^{X}}\left[ S_{%
\pounds /X}(T)\left( K-\omega _{1}S_{\text{\euro /}\pounds }(T)-\omega
_{2}S_{\$/\pounds }(T)\right) _{+}\right]  \label{mf15} \\
&=&\frac{S_{\text{\euro /}\pounds }(0)e^{-r_{\text{\euro }}T}}{\mathbb{E}%
_{Q_{T}^{X}}\left[ S_{_{\text{\euro }}/X}(T)\right] }\mathbb{E}_{Q_{T}^{X}}%
\left[ S_{\pounds /\text{\euro }}^{2/3}(T)S_{\$/\text{\euro }%
}^{-1/3}(T)\left( K-\frac{\omega _{1}}{S_{\pounds /\text{\euro }}(T)}-\omega
_{2}\frac{S_{\$/\text{\euro }}(T)}{S_{\pounds /\text{\euro }}(T)}\right) _{+}%
\right]  \notag \\
&=&S_{\text{\euro /}\pounds }(0)\int_{0}^{\infty }\int_{0}^{\infty }x\left(
K-\frac{\omega _{1}}{x}-\omega _{2}\frac{y}{x}\right) _{+}\frac{\partial ^{2}%
}{\partial x\partial y}U(x,y)dxdy  \notag \\
&=&S_{\text{\euro /}\pounds }(0)\int_{0}^{\infty }\left[ U\left( \infty ,%
\frac{z}{\omega _{2}}\right) -U\left( \frac{z\omega _{2}+\omega _{1}}{K},%
\frac{z}{\omega _{2}}\right) \right] dz,  \notag
\end{eqnarray}%
where 
\begin{equation*}
U(\infty ,K_{2})=e^{-r_{\text{\euro }}T}\left[ N(-d_{2}^{-})+K_{2}\sqrt{T}%
\sigma _{2}^{\prime }(K_{2})N^{\prime }(d_{2}^{-})-1\right] .
\end{equation*}
\end{example}

We note that if we set one of $\omega _{i}$ to zero in (\ref{mf14}), then
the formula gives the EUR price of a put on GBP or USD. Substituting $U$
from (\ref{mf12}) in (\ref{mf14}) with one of $\omega _{i}$ being zero, we
can recover the Black-Sholes price of the corresponding put which means that
the pricing formula (\ref{mf14}) (or what is the same, (\ref{mf68})) exactly
reproduces the plain vanilla data to which the calibration is made. See a
calibration illustration in the next section. The difference with the
approach of \cite{Aus11,Aus14} is that here we obtain the density which can
be used to price on all the three markets based on a single calibration.

\begin{remark}
\label{rem:free}We note that given a smile we get the exact pricing density (%
\ref{mf1}) for a single pair. This allows us to combine marginals for each
pair together with a copula to get a joint density for the triangle instead
of using the approach based on the best-of option as considered above. We do
not explore the use of copulas here, for the corresponding discussion see 
\cite{Aus14}.
\end{remark}

\section{Examples of calibration\label{sec:calib}}

In this section we present calibration examples for the models from Section~%
\ref{sec:hes} and~\ref{sec:esn} and we illustrate the model-free approach of
Section~\ref{sec:mf}. 

We recall \cite{Clark,reiswich2012fx} that the FX market is different to
other financial markets in terms of volatility smile construction and
quoting mechanisms used. FX options are quoted in implied volatility $\sigma 
$, delta $\Delta $ instead of strike $K$, and maturity $T$. The market
convention is to quote three currency pair-specific most commonly traded
options. Their choice depends on a delta hedging and ATM convention \cite%
{reiswich2012fx,dadachanji2016fx}, and typically $25\Delta $ options are
among the considered options. Occasionally, one also uses $10\Delta $
put/call options, as they are widely available but not as liquid as $%
25\Delta $ options \cite{Clark}. The option prices are inverted to calculate
the corresponding volatility values, which are used for constructing the
volatility smile. The data we use in this section for calibration are given
in Table~\ref{tab:exampledata}. 
\begin{table}[tbh]
\caption{FX market data for 1 year maturity options, Bloomberg 03/06/2016.}
\label{tab:exampledata}\centering
\begin{tabular}{|c|c|c|c|}
\hline
& GBP-EUR & USD-EUR & GBP-USD \\ \hline\hline
$\sigma_{25Put\Delta}$ & $12.435\%$ & $9.005\%$ & $11.000\%$ \\ \hline
$\sigma_{ATM}$ & $10.945\%$ & $9.250\%$ & $13.072\%$ \\ \hline
$\sigma_{25Call\Delta}$ & $10.345\%$ & $10.265\%$ & $9.972\%$ \\ \hline
\end{tabular}%
\end{table}

\subsection{Calibration: extended skew normal model}

\label{ssec:calib2d} In this subsection we calibrate the model (\ref%
{eq:distrf}), (\ref{eq:defV}), (\ref{eq:defZ}) from Section~\ref{sec:esn} to
market data for two currency pairs. The use of just three options in
calibration of volatility smiles leads to another typical (and which is in
contrast to other markets) feature of the FX market that the volatility
smile should interpolate the given three data points. Hence FX calibration
is usually done via a root-finding numerical algorithm, while on other
markets, where a large number of option prices are available for
constructing volatility curves, one normally uses least-square type
algorithms for this purpose. 
\begin{figure}[tbh]
\centering
\includegraphics[scale = 0.6]{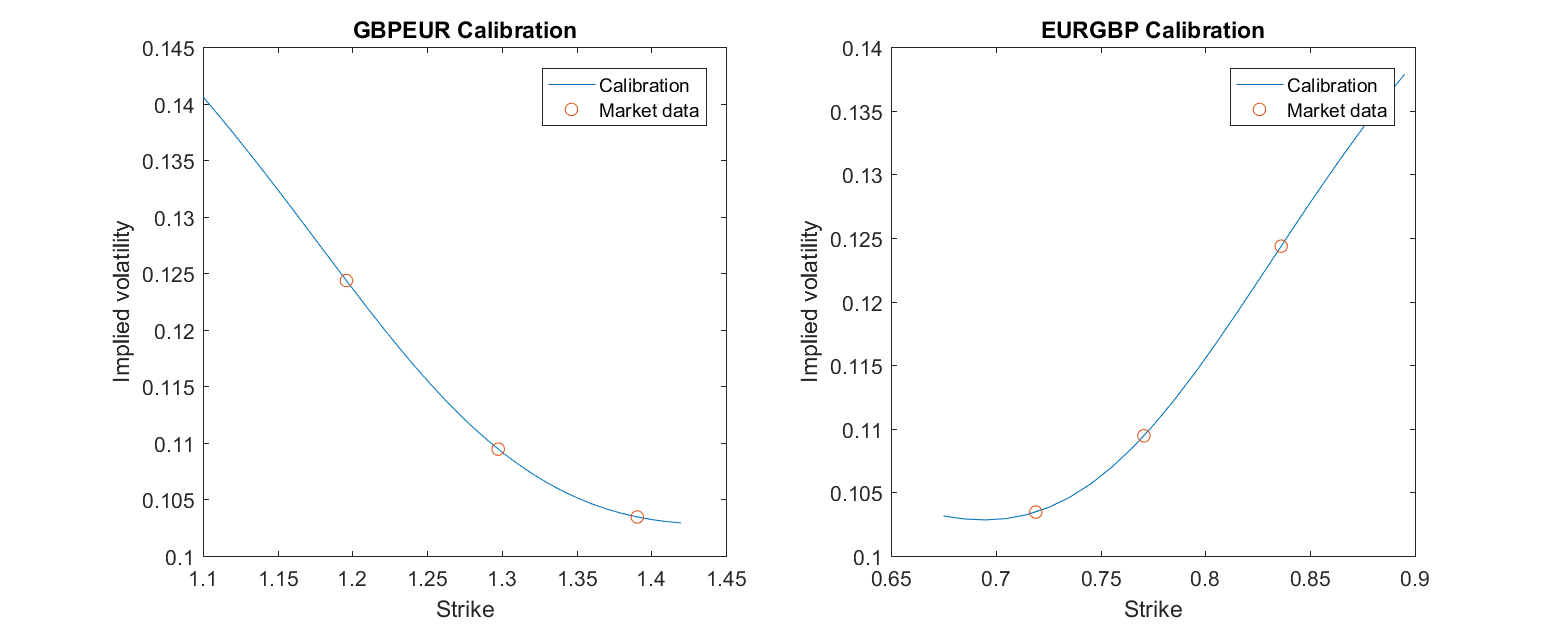}
\caption{Calibration results for the GBP-EUR currency pair (left) and the
inverse pair EUR-GBP (right) with $T=1$, $r_{\text{\pounds }}=0.0025$, $r_{%
\text{\euro }}=0.00$, $S_{\text{\pounds }/\text{\euro }}(0)=1.2935$.}
\label{fig:GBPEUR1}
\end{figure}

\begin{table}[tbh]
\caption{The results of calibration for GBP-EUR and EUR-GBP.}
\label{tab:paramData1}\centering
\begin{tabular}{|c|c|}
\hline
parameter & GBP-EUR/EUR-GBP \\ \hline\hline
$a$ & $0.06297173$ \\ \hline
$\alpha_1$ & $-3.18990817$ \\ \hline
$\alpha_2$ & $1.57557895$ \\ \hline
$\beta_1$ & $-0.5$ \\ \hline
$\beta_2$ & $0.5$ \\ \hline
\end{tabular}
\quad 
\begin{tabular}{|c|c|c|}
\hline
& GBP-EUR & EUR-GBP \\ \hline\hline
$skew$ & $-0.87012308 $ & $0.87012308 $ \\ \hline
$kurtosis$ & $4.94244079 $ & $4.94244079 $ \\ \hline
\end{tabular}%
\end{table}

The calibration was done in MatLab R2016a, where we used the MatLab function 
\textit{fsolve} (which by default uses the built-in \textit{%
trust-region-dogleg} algorithm) to match the option price data (three points
per currency pair). We fixed the (free) parameters $\beta _{1}=-0.5$ and $%
\beta _{2}=0.5$. For the calibration of the GBP-EUR pair, we use $a=\sigma
_{ATM},\alpha _{1}=-3.0$ and $\alpha _{2}=1.0$ as initial values, as the
negative skew of the volatility smile suggests a larger left tail (of the
the distribution of $Z$). The calibration on a standard Desktop computer
(Windows 7, 64-bit, Intel(R) Core(TM) i5-6500 CPU@3.20GHz, 16GB RAM) takes
0.11 seconds.


The calibration results for the GBP-EUR pair are given in Figure~\ref%
{fig:GBPEUR1} and Table~\ref{tab:paramData1}. One can see that the proposed
pricing mechanism (see Theorem~\ref{thm:main} and also (\ref{eq:call4}))
together with the exchange rate model (\ref{eq:distrf}), (\ref{eq:defV}), (%
\ref{eq:defZ}) preserves the volatility smile symmetry as skew, kurtosis
(neglecting natural sign changes) and the model parameters stay the same. We
also confirm that it is sufficient to calibrate the model using the GBP-EUR
data and that the model reproduces both GBP-EUR and EUR-GBP smiles with the
same parameters $a$, $\alpha _{1}$, $\alpha _{2}$, $\beta _{1}$, $\beta _{2}$%
. Moreover, it can been seen that the resulting skew of $~0.87$ and kurtosis
of $~4.94$ indicate the difference of the resulting distribution $Z$ to a
normal distribution ($skew=0$, $kurtosis=3.0$).

The calibration results for the USD-EUR pair are given in Figure \ref%
{fig:USDEUR1} and Table \ref{tab:paramData2}. The same observations as above
for the GBP-EUR pair can be made here as well.

\begin{figure}[htb]
\centering
\includegraphics[scale = 0.6]{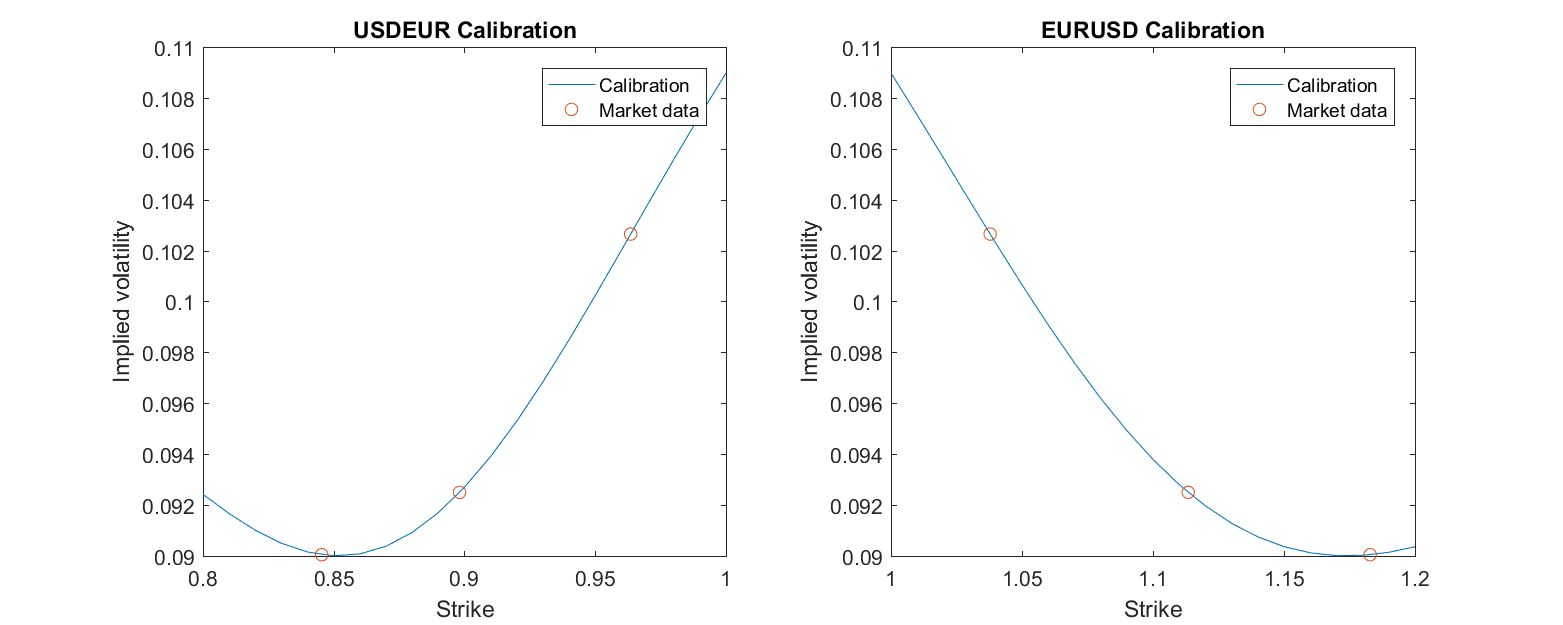}
\caption{Calibration for the USD-EUR currency pair (left) and the inverse
pair EUR-USD (right) with $T=1.0$, $r_{\$} = 0.0025$, $r_{\text{\euro}} =
0.00 $, $S_{\$/\text{\euro}}(0) = 0.8968$.}
\label{fig:USDEUR1}
\end{figure}
\begin{table}[]
\caption{The results of calibration for USD-EUR and EUR-USD.}
\label{tab:paramData2}\centering
\begin{tabular}{|c|c|}
\hline
parameter & USD-EUR/EUR-USD \\ \hline\hline
$a$ & $0.05259980$ \\ \hline
$\alpha_1$ & $-1.94011846$ \\ \hline
$\alpha_2$ & $2.90433341$ \\ \hline
$\beta_1$ & $-0.5$ \\ \hline
$\beta_2$ & $0.5$ \\ \hline
\end{tabular}
\begin{tabular}{|c|c|c|}
\hline
& USD-EUR & EUR-USD \\ \hline\hline
$skew$ & $0.53740761$ & $-0.53740761$ \\ \hline
$kurtosis$ & $4.52666183 $ & $4.52666183 $ \\ \hline
\end{tabular}%
\end{table}

\subsection{Calibration: Heston model}

\label{ssec:calibHeston} In this subsection we calibrate (i.e., find a
parameter set of $\upsilon _{0},\kappa ,\delta ,\theta ,\rho )$ the Heston
model (\ref{hes9}) from Section~\ref{sec:hes} to market data for the GBP-EUR
currency pairs as shown in Table~\ref{tab:exampledata}. Similar to Section~%
\ref{ssec:calib2d}, we used the root-finding method. We fix the parameters
for $\upsilon _{0}$ and $\kappa $ at suitable levels. To compute option
prices, we applied the Monte Carlo technique to the pricing formulas~%
\eqref{hes8} and~\eqref{hes89}. To simulate the Heston model under $Q^{X_{T}}
$~\eqref{hes9}, we used an Euler discretisation scheme for the log forward
prices and a moment-matching scheme for the volatility process \cite%
{rouah2013heston}, which preserves positivity of the volatility process.

\begin{figure}[tbh]
\centering
\includegraphics[scale = 0.45]{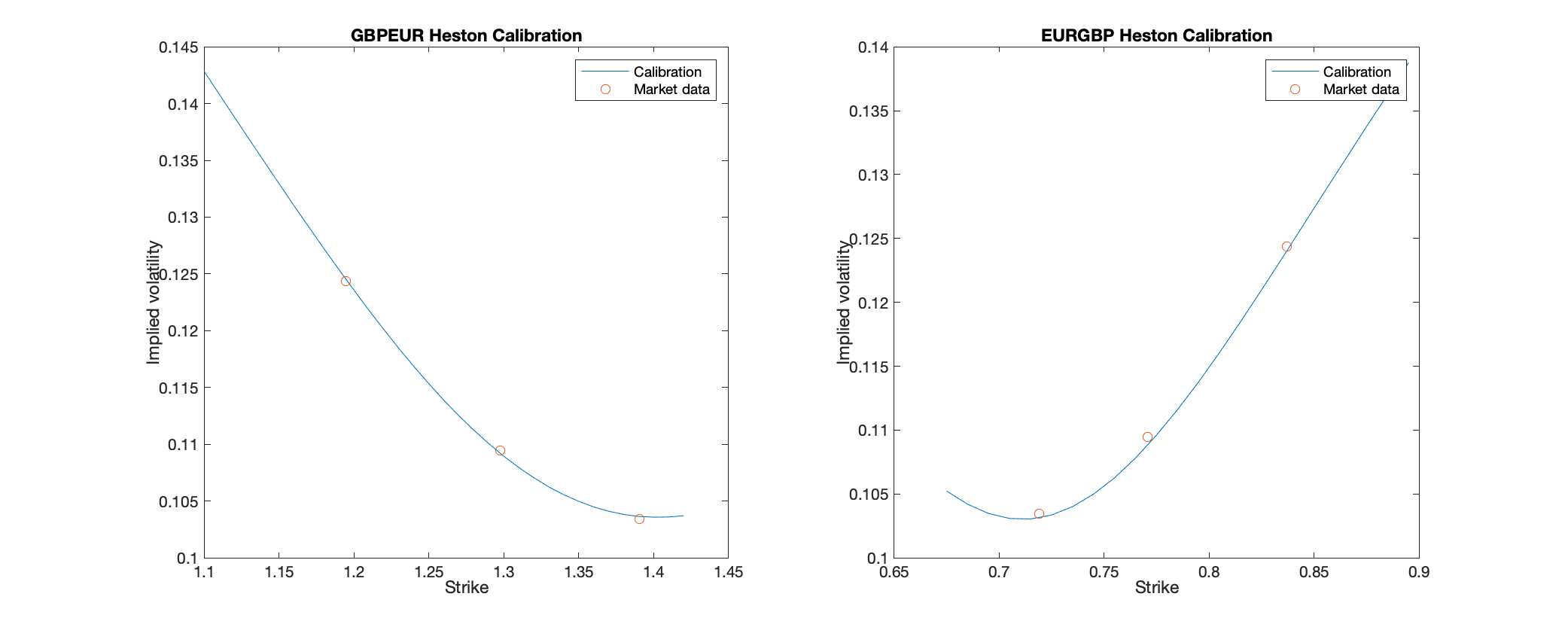}
\caption{Heston model calibration results for the GBP-EUR currency pair
(left) and the inverse pair EUR-GBP (right) with $T=1$, $r_{\text{\pounds }%
}=0.0025$, $r_{\text{\euro }}=0.00$, $S_{\text{\pounds }/\text{\euro }%
}(0)=1.2935$.}
\label{fig:GBPEUR2}
\end{figure}

\begin{table}[tbh]
\caption{The results of Heston model calibration for GBP-EUR and EUR-GBP.}
\label{tab:paramDataHes}\centering
\begin{tabular}{|c|c|}
\hline
parameter & GBP-EUR/EUR-GBP \\ \hline\hline
$N_{MC}$ & $10^9$ \\ \hline
$h$ & $0.05$ \\ \hline
$\upsilon_0$ & $0.0086$ \\ \hline
$\kappa$ & $1.500$ \\ \hline
$\delta$ & $0.71020580946071$ \\ \hline
$\theta$ & $0.02949445852250$ \\ \hline
$\rho$ & $-0.40966532579627$ \\ \hline
\end{tabular}%
\end{table}

The calibration results for the GBP-EUR pair are given in Figure~\ref%
{fig:GBPEUR2} and Table~\ref{tab:paramDataHes}. Again, it can be seen that
the proposed pricing mechanism (see Theorem~\ref{thm:main}) together with
the Heston model \eqref{hes8} is flexible enough to match the GBP-EUR smile.
Moreover, it is sufficient to calibrate the model to the GBP-EUR smile,
which results in the inverse smile EUR-GBP smile to be automatically
calibrated automatically (using the appropriate pricing formula~\eqref{hes89}%
).

\subsection{Illustration of the model-free approach}

\label{ssec:calib3d} In this subsection, we demonstrate how we can
approximate the scaled density function in the model-free approach of
Section~\ref{sec:mf} from market data for three currencies. We recall that
thanks to the intermediate currency approach we can use the same density
function to price options on all three markets. We retrieve the scaled
density by differentiating $U(K_{1},K_{2})$ twice: 
\begin{equation*}
\frac{\partial ^{2}}{\partial K_{1}\partial K_{2}}U(K_{1},K_{2})=\dfrac{%
e^{-r_{\text{\euro }}T}}{\mathbb{E}_{Q_{T}^{X}}\left[ S_{\text{\euro }/X}(T)%
\right] }K_{1}^{-1/3}K_{2}{}^{-1/3}\ \rho (K_{1},K_{2};T).
\end{equation*}%
We use the same market data as before, for the three currency pairs GBP-EUR,
USD-EUR and GBP-USD, which can be found in Table~\ref{tab:exampledata}. We
can find the corresponding strikes by inverting the Garman-Kohlhagen formula
for all three pairs. As we need the volatility smiles to satisfy the growth
condition~\eqref{eq:mf-req1}, we fit a 2nd order polynomial with the three
parameters $p_{i}^{(j)}\in \mathbb{R},\quad j=1,2,3,$ to the implied
volatility data transformed by $\exp \left[ \sigma _{i}^{2}(K)\right] .$
Then we obtain the interpolated implied volatilities as 
\begin{equation*}
\tilde{\sigma}_{i}(K)=\sqrt{\log \left[
p_{i}^{(1)}K^{2}+p_{i}^{(2)}K+p_{i}^{(3)}\right] }.
\end{equation*}%
\begin{figure}[tbp]
\centering
\includegraphics[scale = 0.6]{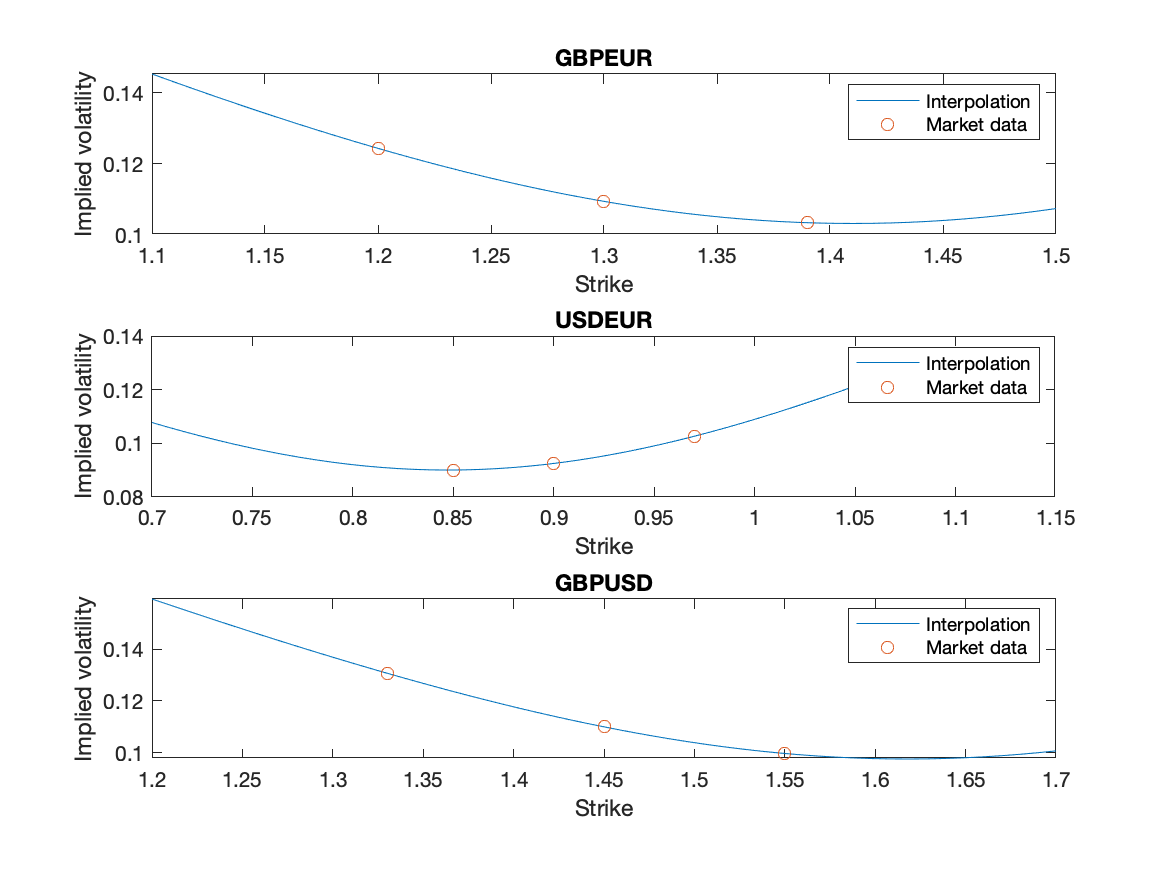}
\caption{Implied volatility interpolation for GBP-EUR, USD-EUR and GBP-USD
pairs with $T=1.0$, $r_{\$}=0.0025$,$r_{\$}=0.0025$, $r_{\text{\euro }}=0.00$%
, $r_{\text{\pounds }}=0.0025$, $r_{\text{\pounds }}=0.0025$, $S_{\$/\text{%
\euro }}(0)=0.8968$, $S_{\text{\pounds }/\text{\euro }}(0)=1.2935$, $S_{%
\text{\pounds }/\$}=1.4423$.}
\label{fig:mf-3D-smiles}
\end{figure}
The results of the interpolation for the implied volatility smiles can be
seen in Figure~\ref{fig:mf-3D-smiles}. The partial derivative with respect
to $K_{1}$ and $K_{2}$ of $U(K_{1},K_{2})$ can be found by numerically
differentiating~\eqref{mf12} on a fine grid of $K_{1}$ and $K_{2}$. We use
the MATLAB function \textit{diff} to compute the point-wise $\frac{\partial
^{2}}{\partial K_{1}\partial K_{2}}U(K_{1},K_{2})$ surface for a range of
strikes $K_{1}$ and $K_{2}$. Note that $K_{3}=\frac{K_{1}}{K_{2}}$. The
resulting surface and contour plots are given in Figure~\ref%
{fig:mf-surface-contour-plot}. We remark that $\frac{\partial ^{2}}{\partial
K_{1}\partial K_{2}}U(K_{1},K_{2})$ is positive for the whole range of
strikes considered as required. 
\begin{figure}[tbp]
\centering
\begin{subfigure}{.5\textwidth}
  \centering
  \includegraphics[scale = 0.45]{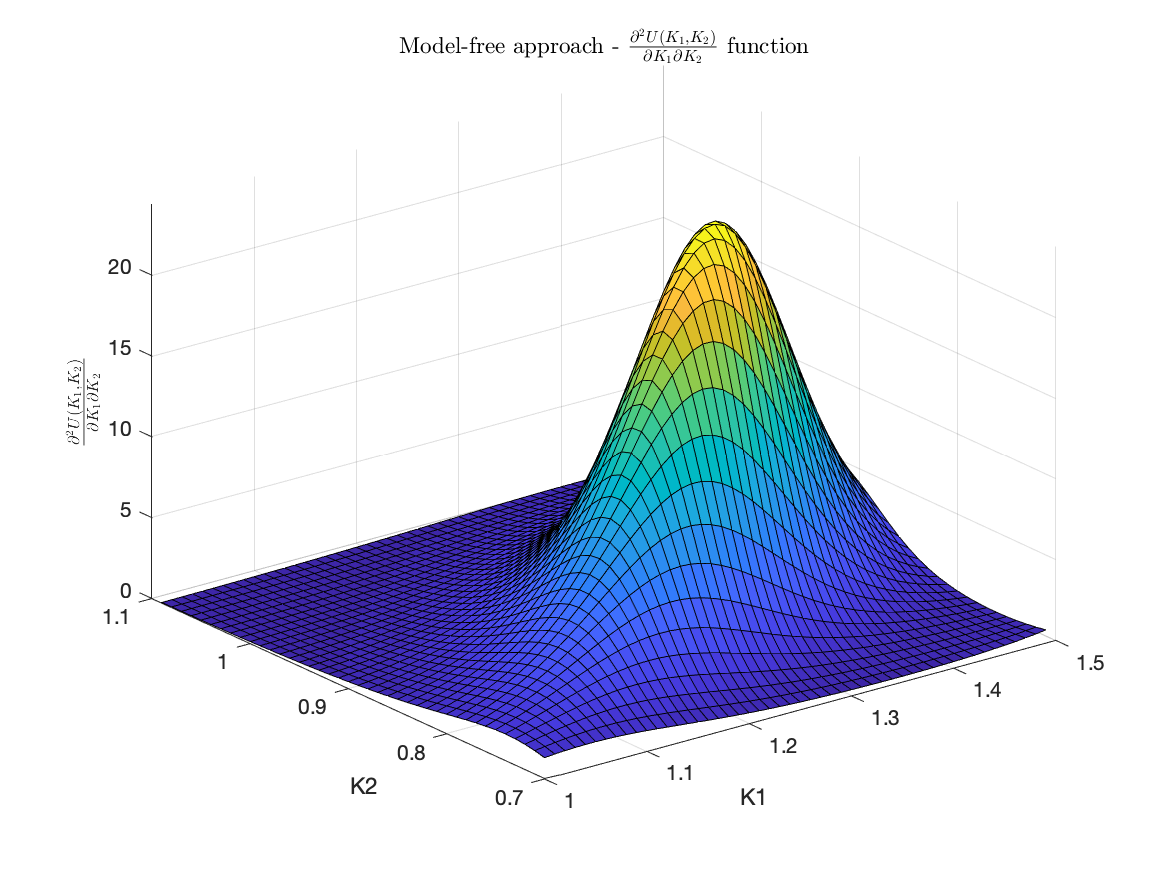}
\end{subfigure}%
\begin{subfigure}{.5\textwidth}
  \centering
  \includegraphics[scale = 0.45]{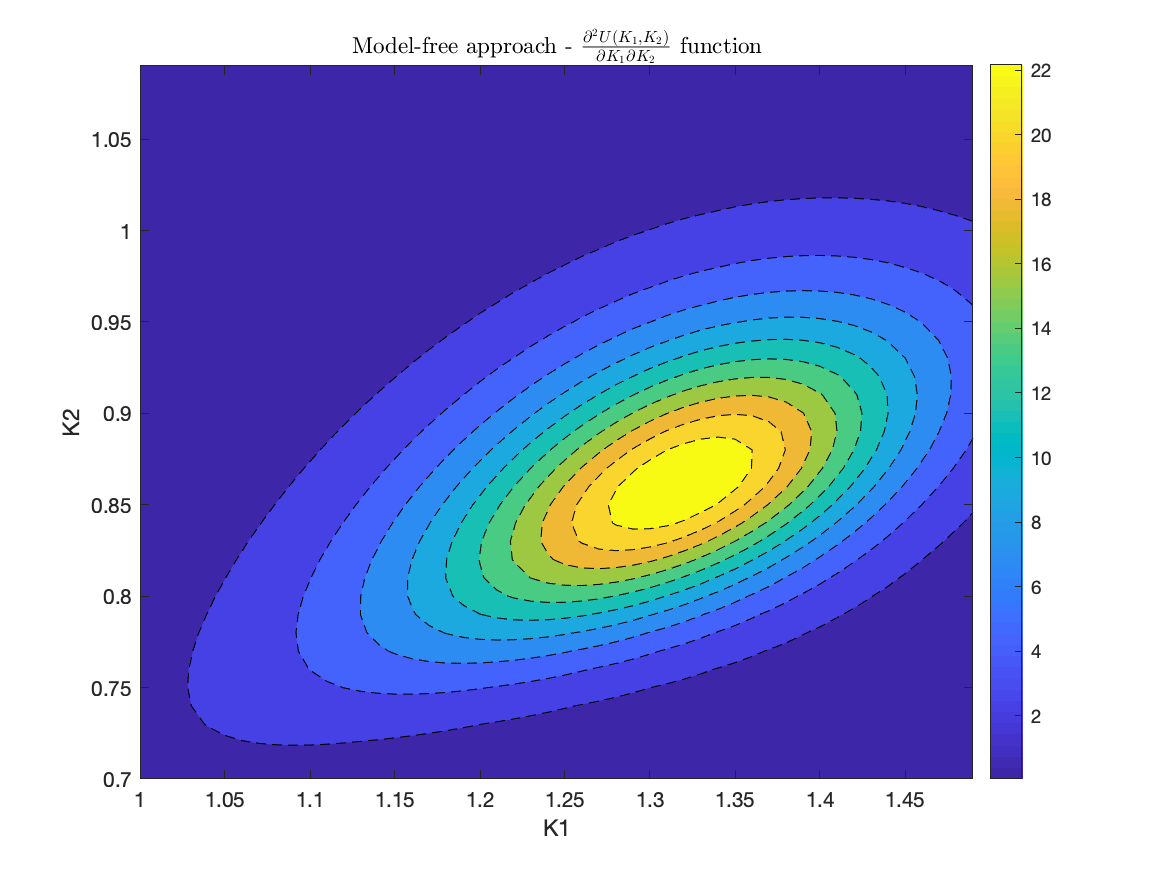}
\end{subfigure}
\caption{Implied scaled density surface and contour plot for the three
currency pairs for a range of strikes $K_{1}$ and $K_{2}$.}
\label{fig:mf-surface-contour-plot}
\end{figure}

\bibliographystyle{plain}
\bibliography{literature}

\end{document}